\documentclass[useAMS,usenatbib,aas_macros]{mnras}
\usepackage{graphicx}
\usepackage{amssymb}
\usepackage{footnote}

\usepackage{epsfig}
\usepackage{color}
\usepackage{amssymb}
\usepackage{ams math}
\usepackage[english]{babel}
\usepackage{subfig}
\usepackage{relsize}
\usepackage{xfrac}

\def\hii{\mbox{H\,{\sc ii}}}

\def\lir{L$_{\rm{IR}}$}

\title[FiBY: dust attenuation at $z\simeq5$]{The First Billion Years project: constraining the dust attenuation law of star-forming galaxies at $\bmath{z\simeq5}$}
\author[F. Cullen]{\parbox\textwidth{F. Cullen$^{1}$\thanks{E-mail:fc@roe.ac.uk}, 
R.J. McLure${^{1}}$,
S. Khochfar${^{1}}$,
J.S. Dunlop${^{1}}$,
C. Dalla Vecchia${^{2,3}}$}\\\\
$^{1}$SUPA\thanks{Scottish Universities Physics Alliance}, Institute for Astronomy, University of Edinburgh, Royal Observatory, Edinburgh EH9 3HJ\\
$^{2}$Instituto de Astrof\'isica de Canarias, C/V\'ia L\'actea s/n, E-38205 La Laguna, Tenerife, Spain\\
$^{3}$Departamento de Astrof\'isica, Universidad de La Laguna, Av. del Astrof\'isico Franciso S\'anchex s/n, E-38206 La Laguna, Tenerife, Spain}

\begin{document}

\date{Accepted -- . Received 2016 May 30}

\pagerange{\pageref{firstpage}--\pageref{lastpage}} \pubyear{2016}

\maketitle	

\label{firstpage}

%%%%%%%%%%%%%%%%%% ABSTRACT %%%%%%%%%%%%%%%%%%%
\begin{abstract}
We present the results of a study investigating the dust attenuation law at $z\simeq 5$, based on synthetic spectral energy distributions (SEDs) calculated for a sample of N=498 galaxies drawn from the First Billion Years (FiBY) simulation project. 
The simulated galaxies at $z\simeq 5$, which have M$_{1500} \leq -18.0$ and $7.5 \leq \rm{log(M/M}_{\odot}\rm{)} \leq 10.2$, display a mass-dependent $\alpha$-enhancement, with a median value of $[\alpha/\rm{Fe}]_{z=5}~\simeq~4~\times~[\alpha/\rm{Fe}]_{Z_{\odot}}$. 
The median Fe/H ratio of the simulated galaxies is $0.14\pm0.05$ which produces steep intrinsic UV continuum slopes; $\langle \beta_{i} \rangle = -2.4 \pm 0.05$. 
Using a set of simple dust attenuation models, in which the wavelength-dependent attenuation is assumed to be of the form $A(\lambda) \propto \lambda^{n}$, we explore the parameter values which best reproduce the observed $z=5$ luminosity function (LF) and colour-magnitude relation (CMR). 
We find that a simple model in which the absolute UV attenuation is a linearly increasing function of log stellar mass ($\rm{A}_{1500}=0.5\times \rm{log(M/M_{\odot})} - 3.3$), and the dust attenuation slope ($n$) is within the range $-0.7 \leq n \leq-0.3$, can successfully reproduce the LF and CMR over a wide range of stellar population synthesis model (SPS) assumptions, including the effects of massive binaries. 
This range of attenuation curves is consistent with a power-law fit to the Calzetti attenuation law in the UV ($n=-0.55$).
In contrast, curves as steep as the Small Magellanic Cloud (SMC) extinction curve ($n=-1.24$) are formally ruled out.
Finally, we show that our models are consistent with recent 1.3mm ALMA observations of the Hubble Ultra Deep Field (HUDF), and predict the form of the $z\simeq5$ IRX$-\beta$ relation.
\end{abstract}


%%%%%%%%%%%%%%%%% INTRODUCTION %%%%%%%%%%%%%%%%%

\section{Introduction}

Tracing the star-formation rate of galaxies across cosmic time, and particularly at $z \gtrsim 3$ (prior to the peak epoch of cosmic star formation), remains one of the most important focuses of observational cosmology, key to shedding light on the build up of stellar mass and metal content at early times, and clarifying the process of cosmic reionization.
However, despite the substantial progress that has been made in constraining the star-formation rate density (SFRD) out to $z\approx8$ \citep[e.g.][]{madau2014}, the lack of far-infrared (FIR) data for galaxies at $z \gtrsim 3$ means that, at these redshifts, star-formation rate estimates are mainly derived solely from rest-frame ultraviolet (UV) to optical observations.
Therefore, our knowledge of the SFRD at $z \gtrsim 3$ is affected by inherent uncertainties regarding the correct dust corrections to apply at high redshifts. 

When only rest-frame UV data are available, the observed UV luminosities must be dust-corrected based on correlations derived at lower redshifts.
Most commonly, the \citet{meurer1999} (M99) IRX-$\beta$ relation (IRX=$\sfrac{L_{IR}}{L_{UV}}$) is used to estimate the intrinsic UV luminosity from the observed UV continuum slope $\beta$ ($f_{\lambda} \propto \lambda^{\beta}$).
This relation was derived from a sample of local starburst galaxies and has been shown to apply for star-forming galaxies out to $z \simeq 2$ \citep[e.g.][]{reddy2010, reddy2012}.
Under the assumption that the dust heating is dominated by a young stellar population, IRX can be related directly to the attenuation in the UV \citep[A$_{UV}$, see][]{meurer1999}, allowing one to relate the absolute UV attenuation to the observed shape of the UV continuum slope, assuming the intrinsic shape is known.
The form of the subsequent A$_{UV}$-$\beta$ relation measured for local starbursts is what one would expect for a relatively grey reddening law, similar to that derived by \citet{calzetti2000}.
Steeper laws such as the Small Magellanic Cloud (SMC) extinction curve \citep[e.g.][]{gordon2003}, or laws which include a strong $2175 \rm{\AA}$ absorption feature \citep[e.g., Milky Way;][]{cardelli1989} are not compatible with the local starburst galaxies.
Therefore, when applying the \citet{meurer1999} IRX-$\beta$ relation at $z \gtrsim 3$, one is implicitly assuming Calzetti-like reddening.
Furthermore, in cases where rest-frame UV plus optical data is available, the intrinsic UV luminosities are typically derived via detailed SED-fitting \citep[e.g.][]{mclure2011, bowler2015}, and the subsequent UV dust-correction depends on which reddening laws are assumed for the fitting.
Again, and motivated in part by its compatibility with the local IRX-$\beta$ relation, by far the most common assumption is the \citet{calzetti2000} reddening law.
Thus, this assumption is implicit in our current understanding of the UV luminosity function and SFRD evolution at high redshifts. 

However, with the advent of the Atacama Large Millimeter Array (ALMA), the form of the attenuation law at high redshifts has begun to be questioned.
Recent ALMA observations have indicated that the IR luminosities of `typical' star-forming galaxies at $z \simeq 3$ are lower than expected assuming a Calzetti law, and have been interpreted as evidence that the attenuation curve is as steep, or steeper, than the SMC extinction curve at these redshifts \citep[][]{capak2015, bouwens2016}.
Similar results have also previously been reported for young galaxies ($<$ 100 Myr old) \citep[e.g.][]{reddy2010, reddy2012} and sample of gravitationally lensed galaxies \citep[e.g.][]{baker2001, siana2008, siana2009} at these redshifts.
For a steeper, SMC-like, law a given amount of reddening in the UV is achieved with less absolute attenuation (i.e. the $A_{UV}-\beta$ relation becomes much shallower).
If dust reddening at high redshift is SMC-like, then previous UV luminosity corrections made assuming the \citet{meurer1999} IRX-$\beta$ relation, or \citet{calzetti2000}, will have overestimated the dust attenuation, and hence overestimated the intrinsic luminosities and SFRs at $z \simeq 3$.
On the other hand, direct measurements of the attenuation curve at high redshifts have yet to find strong evidence for the steep SMC-like curves inferred from IR measurements.
For example, \citet{reddy2015} found a Calzetti-like attenuation curve shape at $\lambda<2600\rm{\AA}$ for a sample of N=224 star-forming galaxies at $1.4 < z < 2.6$.
Using an independent method, \citet{zeimann2015_dust} found an average attenuation curve at $1.90<z<2.35$ similar to the Calzetti curve, and perhaps even greyer at lower masses, with a tentative detection of a mass dependence of the attenuation curve slope.
Furthermore, \citet{scoville2015} performed a similar analysis with a sample of N=266 galaxies at $2.0 < z < 6.5$ and again found no evidence for a deviation from a Calzetti-like attenuation curve.
Interestingly, the \citet{scoville2015} analysis is the only example of a direct attenuation curve measurement at the same redshifts probed by recent ALMA observations, and does not support the existence of a steep SMC-like curve.

In light of these recent results, the primary motivation of this paper is to use state-of-the-art hydrodynamical simulations of galaxies at $z=5$, from the First Billion Years project (FiBY) \citep[][Khochfar et al. in prep]{paardekooper2013}, to constrain the form of the attenuation curve, by comparing the SEDs of simulated galaxies to the luminosity function (LF) and colour magnitude relation (MR) at $z=5$.
This specific redshift was chosen primarily because it is the highest redshift at which there is good consistency between measurements of the LF and CMR across a variety of independent studies \citep[e.g][]{bouwens2012,rogers2014,bowler2015,bouwens2015}; furthermore, it is at this approximate redshift that a number of recent ALMA observations have been made claiming evidence for an evolution of dust properties in the early Universe.

The structure of this paper is as follows: in Section \ref{sec_sed_gen} we describe how we extracted the relevant stellar data for galaxies in our sample from the FiBY simulation, and outline the method we employed to generate synthetic SEDs, including a description of SPS models and photoionization model assumptions. In Section \ref{sec_beta_m1500} we review the intrinsic properties of our synthetic SEDs and describe how these properties are affected by our stellar population synthesis (SPS) and photoionization model choices. In Section \ref{sec_dust_model} we describe the two dust models we have employed to constrain various properties of the attenuation law, before presenting the results of fitting our sample of simulated galaxies to the observed LF and CMR in Section \ref{sec_fitting}. In Section \ref{sec_IR_predictions} we discuss the IR properties of $z=5$ galaxies predicted by our models and compare these to recent ALMA observations. Finally, in Sections \ref{sec_discussion} and \ref{sec_conclusions} we provide a discussion and summary of our main conclusions.
The FiBY simulation adopts the following cosmological parameters, that we will also adopt throughout this paper: $\Omega_n =0.265$, $\Omega_b = 0.0448$, $\Omega_\Lambda =0.735$, $H_0 =71$ km s$^{-1}$ Mpc$^{-1}$ and $\sigma_8=0.81$.

\section{Generating Synthetic SEDs}\label{sec_sed_gen}

In this section we discuss how the synthetic SEDs were generated from the FiBY simulation data, including a detailed discussion of our SPS and photoionization modelling assumptions, and an overview of the physical properties of our final simulated galaxy sample.

\subsection{SPS models}\label{sec_sps}

To construct the SED for each galaxy we used the latest release of `Binary Population and Spectral Synthesis' \citep[BPASSv2;][]{eldridge2016,stanway2016}.
The BPASSv2 models were chosen because, along with conventional single-star models, they also incorporate the evolution of massive stars in binary systems.
Observations in the local Universe have shown that a potentially substantial fraction ($\gtrsim 70\%$) of massive stars undergo a binary interaction (such as mass transfer or a merger) during their lifetimes \citep[e.g.][]{sana2012}.
Moreover, recent studies of galaxies at $z \simeq 2-3$ indicate that the BPASSv2 models are best able to replicate the observed nebular and stellar properties of star-forming galaxies at these redshifts \citep{steidel2016, strom2017}.
We have checked that, for the purposes of this work, BPASSv2 single star models are equivalent to the more commonly adopted STARBURST99 models \citep{leitherer1999}.
In the context of this paper, the main effect of including massive binary evolution in SPS models is to generate harder ionizing spectra, particularly at low metallicities, and to boost UV output at older stellar ages relative to single-star populations.

We considered four sets of models which we refer to according to the upper mass cutoff of the initial mass function (IMF) and whether or not binary evolution is included.
Each galaxy is assigned four separate SEDs corresponding to each of the four models.
BPASSv2-100bin models include binary star evolution with an IMF cutoff of 100 M$_{\odot}$, and BPASSv2-100 are the equivalent single-star evolution models; similarly, BPASSv2-300bin models include binary evolution with an IMF cutoff of 300 M$_{\odot}$, and BPASSv2-300 are the equivalent single-star evolution models.
All models have an IMF index of $-1.3$ between $0.1 - 0.5$ M$_{\odot}$ and $-2.35$ above 0.5 M$_{\odot}$ and cover the following metallicities: Z$_{*}$ = 0.001, 0.002, 0.003, 0.004, 0.006, 0.008, 0.010, 0.014, 0.020, 0.030, 0.040.

\subsection{FiBY simulation data}\label{sec_fiby_data}

Here, we give a brief summary of the FiBY simulations but refer the reader to other papers for a more detailed description \citep[e.g.][]{johnson2013, paardekooper2015}. 
The FiBY simulation suite is a set of high-resolution cosmological hydrodynamics simulations using a modified version of the GADGET code used in the Overwhelmingly Large Simulations (OWLS) project \citep{schaye2010}. 
These simulations reproduce the stellar mass function and star formation rate of galaxies at $z\geq 6$ (Khochfar et al. in prep.), and at the same time also recover the right trends in the metallicity evolution of galaxies (Dalla Vecchia et al. in prep.). 
Furthermore, the simulations recover that low mass galaxies can reionize the Universe \citep{paardekooper2013, paardekooper2015}, and that supernovae-feedback can drive the formation of cores in galaxies \citep{davis2014}.

The code tracks metal pollution for 11 elements: H, He, C, N, O, Ne, Mg, Si, S, Ca and Fe and calculates the cooling of gas based on line-cooling in photoionization equilibrium for these elements \citep{wiersma2009} using tables pre-calculated with CLOUDY v07.02 \citep{ferland1998}. 
Furthermore, the simulation incorporates full non-equilibrium primordial chemistry networks \citep{abel1997,galli1998,yoshida2006} including molecular cooling functions for H$_2$ and HD. 
Star formation is modelled using the pressure law implementation of \citet{schaye2008}, which yields results consistent with the Schmidt-Kennicutt law \citep{schmidt1959, kennicutt1998_schmidt}. 
The threshold density above which stars form is set to $n = 10$ cm$^{-3}$. 
The simulations include feedback from stars is by injecting thermal energy in the neighboring particles \citep{dalla_vecchia2012}. 
For each Pop II supernova $10^{51}$ erg is injected once a star particle has reached an age of 30 Myr, corresponding to the maximum lifetime of stars that end their lives as core-collapse supernovae. 

In this work we will focus on the FiBY\_L  and FiBY\_XL simulations which cover co-moving volumes of (16 Mpc)$^3$ and (32 Mpc)$^3$ respectively ($\approx 3.7 \times 10^4$ Mpc$^3$ combined comoving volume). 
The individual gas and star particle masses in the simulations are log(M/M$_{\odot}$) = 5.68 for the FiBY\_XL  and 4.81 for the FiBY\_L. 
To build the spectral energy distributions of galaxies at $z=5$, halos at this redshift were identified in the simulation using the SUBFIND algorithm \citep{springel2001, dolag2009}.
Then, for each halo, we extracted the masses, ages and metallicities of each star particle.
The mass of each star particle is defined as the total initial mass as opposed to the current mass (i.e. at $z=5$). since the effects of mass-loss with stellar age are accounted for in the BPASSv2 stellar population synthesis models we use to generate the SEDs.
The age of each star particle was calculated using the formation redshift and assumed cosmological parameters.
Finally, we extracted the mass in metals for each star particle; for reasons which will be described in Section \ref{sec_fe_alpha_metal}, we tracked both the total mass in metals and the mass in Fe.
To ease computation time, we binned star particles with ages $>20$ Myr into bins of width 10 Myr, taking averages of mass, age and metallicity for each bin; star particles with ages $\leq 20$ Myr were stored individually.

To build a galaxy SED from a given model we looped over each star particle associated with the galaxy and assigned the stellar spectral energy distribution which best matched the age and metallicity of the particle.
Since the BPASSv2 stellar SEDs are modelled as an instantaneous starburst with a total mass of 1 x 10$^{6}$ M$_{\odot}$, we additionally scaled the flux to the mass of the star particle.
We constructed the final galaxy SED by summing the SEDs of all the individual star particles. 

For our final sample, we selected halos at $z=5$ with an absolute UV magnitude magnitude $\rm{M}_{1500} \leq -18.0$ where $\rm{M}_{1500}$ was calculated by integrating the rest-frame BPASSv2-100bin SED within a top-hat filter of width 100 $\rm{\AA}$ at a central wavelength of $1500 \rm{\AA}$.
This magnitude limit differs slightly according to the assumed SPS model, the limiting value of $\rm{M}_{1500} = -18.0$ for the BPASSv2-100bin model corresponds to $-17.89$, $-18.08$ and $-17.95$ for the BPASSv2-100, BPASSv2-300bin and BPASSv2-300 models respectively.
However, changing the SPS model for which the magnitude cut is defined does not have a significant effect on the physical properties of our sample, or any of the results presented in this paper.

\begin{figure}
        \centerline{\includegraphics[width=\columnwidth]{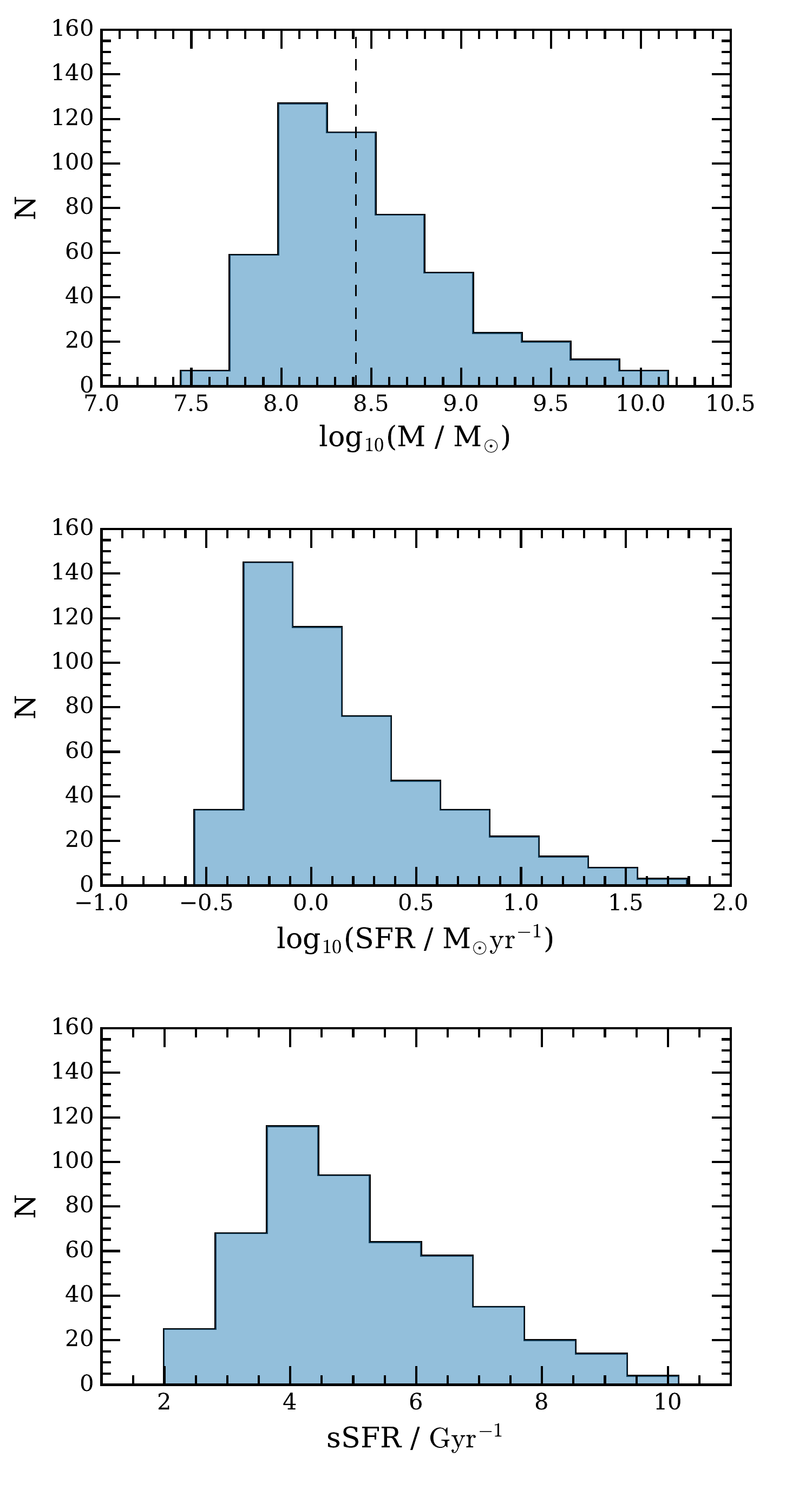}}
        \caption{Stellar mass, star-formation rate and specific star-formation rate distributions for the N=498 galaxies in our simulated $z=5$ sample. The mass is the `current' (i.e. $z=5$) mass in stars, and the SFR is averaged over the last 100 Myr. The median stellar mass is 2.3 x 10$^8$ M$_{\odot}$, the median SFR is 1.1 M$_{\odot}\rm{yr}^{-1}$ and the median sSFR is 4.8 Gyr$^{-1}$.
        The vertical dashed line in the upper panel indicates the mass completeness of the sample: 2.6 x 10$^8$ M$_{\odot}$.} 
        \label{fig_FiBY_galaxy_properties}
    \end{figure}

The final sample contains N=498 galaxies; the distributions of current mass (i.e. at $z=5$), star-formation rate (SFR) (averaged over the past 100 Myr), and specific star-formation rate (sSFR $\equiv$ SFR / M$_{*}$) of our $z=5$ sample are shown in Fig. \ref{fig_FiBY_galaxy_properties}.
The maximum galaxy mass in our sample is 1.4 x 10$^{10}$ M$_{\odot}$, with a median value of 2.3 x 10$^8$ M$_{\odot}$; the sample is mass complete down to 2.6 x 10$^8$ M$_{\odot}$.
The star-formation rates range from 0.3 - 60 M$_{\odot}\rm{yr}^{-1}$ with a median of 1.1 M$_{\odot}\rm{yr}^{-1}$.
The median sSFR is 4.8 Gyr$^{-1}$.

\subsubsection{Stellar metallicity}\label{sec_fe_alpha_metal}

    \begin{figure*}
        \centerline{\includegraphics[width=7.0in]{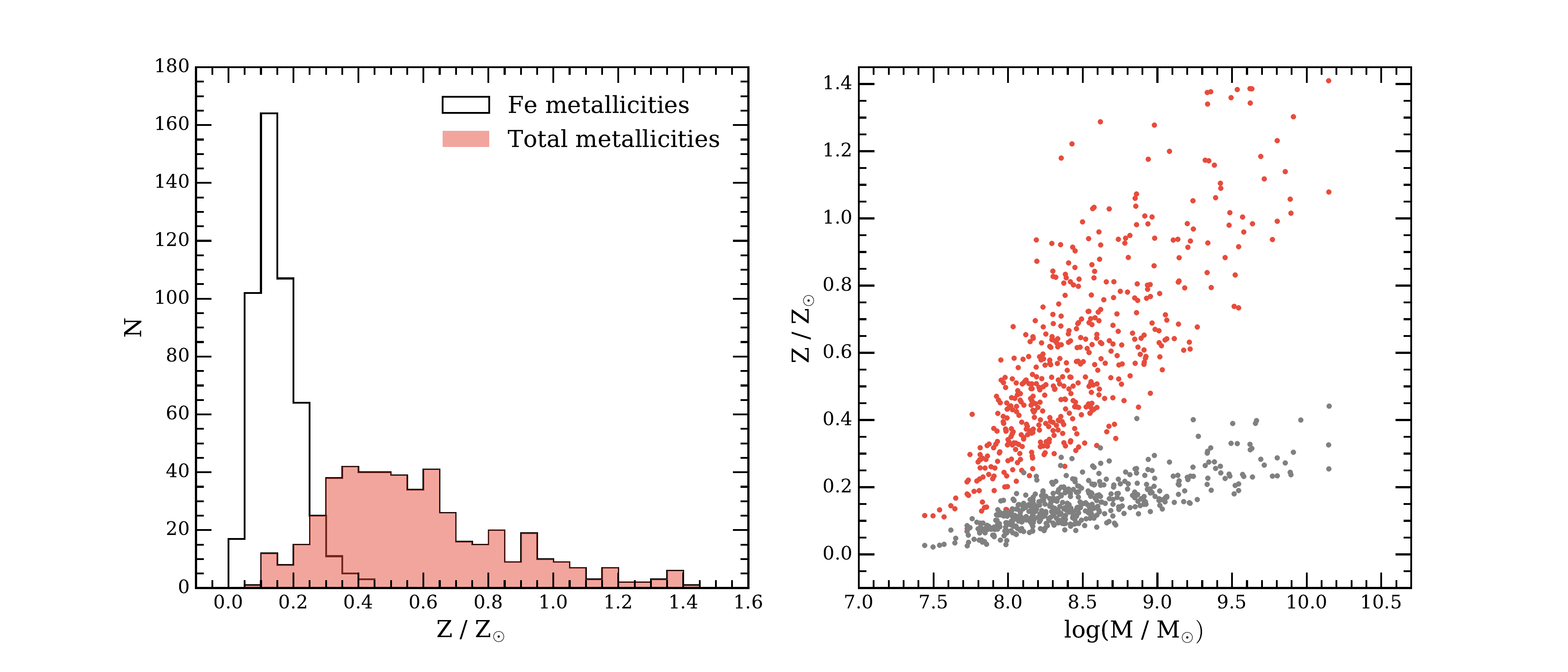}}
        \caption{\emph{Left panel:} Distributions of the total mass in metals, and total mass in Fe (relative to solar) for the N=498 galaxies in our $z=5$ sample weighted by UV luminosity. The median total metallicity is 0.53 Z$_{\rm{Tot}}$/Z$_{\odot}$ and the median Fe metallicity is 0.14 Z$_{\rm{Fe}}$/Z$_{\odot}$. \emph{Right panel:} Relationship between metallicity, both Fe (grey points) and total (red points), and stellar mass.} 
        \label{fig_zFe_zAlpha}
    \end{figure*}

The BPASSv2 models assume solar element abundance ratios, however it is not obvious that the abundance ratios at $z=5$ should be solar; in particular, given the typical star-formation histories at $z=5$, there could be differences in the Fe/H and $\alpha$/H ratios relative to solar values.

The star-formation histories of the galaxies in the FiBY simulation are typically rising (Khochfar et. al. in prep), consistent with other observations and simulations at $z>2$ \citep[e.g.][]{papovich2011, finlator2011, reddy2012b}.
For rising (or constant) star-formation histories the rate of enrichment of $\alpha$ elements from Type II (core-collapse) supernovae, which has a typical time scale of $\sim$ 10 Myr, will exceed that rate of Fe enrichment, since Fe is mainly returned to the ISM via SNe Ia, which have a typical timescale of $\sim$ 1 Gyr.
Therefore, rapidly star-forming galaxies are expected to have enhanced $\alpha/\rm{Fe}$ ratios relative to the solar value.
Indeed, it is not strictly necessary to appeal to star-formation histories to understand this offset at $z=5$, since the median age of the oldest star particle across our galaxy sample is 0.74 Gyr, with a maximum of 0.94 Gyr, implying that the majority of galaxies have not had time to undergo significant SNIa enrichment, regardless of their star-formation histories.
As we discuss below, this $\alpha/\rm{Fe}$ enhancement can have a significant effect on the shape of the UV spectrum.

We were able to calculate the detailed abundance ratios of all galaxies in our sample because the mass fractions for 11 individual elements (H, He, C, N, O, Ne, Mg, Si, S, Ca and Fe) are tracked in the FiBY simulations, along with the total mass fraction in all metals.
Fig. \ref{fig_zFe_zAlpha} shows the distributions of UV-weighted Fe metallicity and total metallicity of the star particles (hereafter referred to, for simplicity, as the $\alpha$ element metallicity), relative to solar, for the N=498 galaxies in our sample, where the solar mass fractions have been taken from \citet{asplund2009}. 
Consistent with the rising star-formation histories, the distributions in Fig. \ref{fig_zFe_zAlpha} (left-hand panel) show the difference between the $\alpha$ and Fe-based metallicities.
The median $\alpha/\rm{Fe}$ ratio in our sample is $\simeq$ 3.9 times larger than the solar value, with the median Fe and $\alpha$ metallicities being Z$_{\rm{Fe}}$/Z$_{\odot}$ = 0.14 and Z$_{\rm{\alpha}}$/Z$_{\odot}$ = 0.53 respectively.
The right-hand panel of Fig. \ref{fig_zFe_zAlpha} shows that, under both definitions, metallicity is an increasing function of stellar mass, however Z$_{\rm{\alpha}}$/Z$_{\odot}$ shows a stronger evolution, implying that the $\alpha/$Fe ratio is also an increasing function of stellar mass.

Non-solar $\alpha/$Fe ratios have an important effect on the details of the emergent SED.
Recently, \citet{steidel2016} performed detailed stellar population synthesis and photoionization modelling of the composite rest-frame UV and optical spectra of 30 star-forming galaxies at $z=2.4$ with log(M/M$_{\odot}$)$>9.0$.
They found that simultaneously fitting (i) the shape of the stellar spectrum in the UV, and (ii) the observed UV + optical emission line ratios, implied a difference of a factor $\simeq 4 - 5$ between the stellar and nebular metallicities (Z$_{*}$/Z$_{\odot} \simeq 0.1$ and Z$_{\rm{neb}}$/Z$_{\odot} \simeq 0.4 - 0.5$).
This is consistent with the offset in our sample (Fig. \ref{fig_zFe_zAlpha}); in fact, restricting our sample to galaxies above 10$^9$ M$_{\odot}$ yields a median $\alpha/\rm{Fe}$ ratio of 4.2.
As argued by \citet{steidel2016}, this offset is naturally explained if star-formation histories at these redshifts are typically rising \citep[arguments for rising SFHs at these redshifts have been made in a number of studies e.g.][]{papovich2011, reddy2012}, and by the fact that stellar opacity in the UV is dominated by Fe, whereas the nebular metallicity is indicative of the abundances of the important coolants in ionized gas (e.g O, N, C), and hence traces the $\alpha$ elements.
To a first approximation, any differences between stellar and nebular metallicities is therefore plausibly a result of the same $\alpha$/Fe offset illustrated in Fig. \ref{fig_zFe_zAlpha}; the lower stellar metallicity is driven by the sensitivity of the stellar UV spectrum to Fe/H, whereas the nebular metallicity follows $\alpha$/H.
Taken together, all these considerations imply that, when applying SPS models with solar abundance sets to galaxies in which one suspects there will be an enhanced $\alpha/\rm{Fe}$ ratio, the stellar metallicity should be defined as Fe/H, rather than the mass fraction of all metals as is commonly assumed, at least when studying the properties of the UV spectrum.
Ideally, one would use SSPs with non-solar abundance sets however, to our knowledge, there is no publicly available SPS code which accounts for both binary star evolution and also allows non-solar element abundances.
As emphasized by \citet{steidel2016}, this should be a key goal for any future SPS models.

\begin{figure}
        \centerline{\includegraphics[width=\columnwidth]{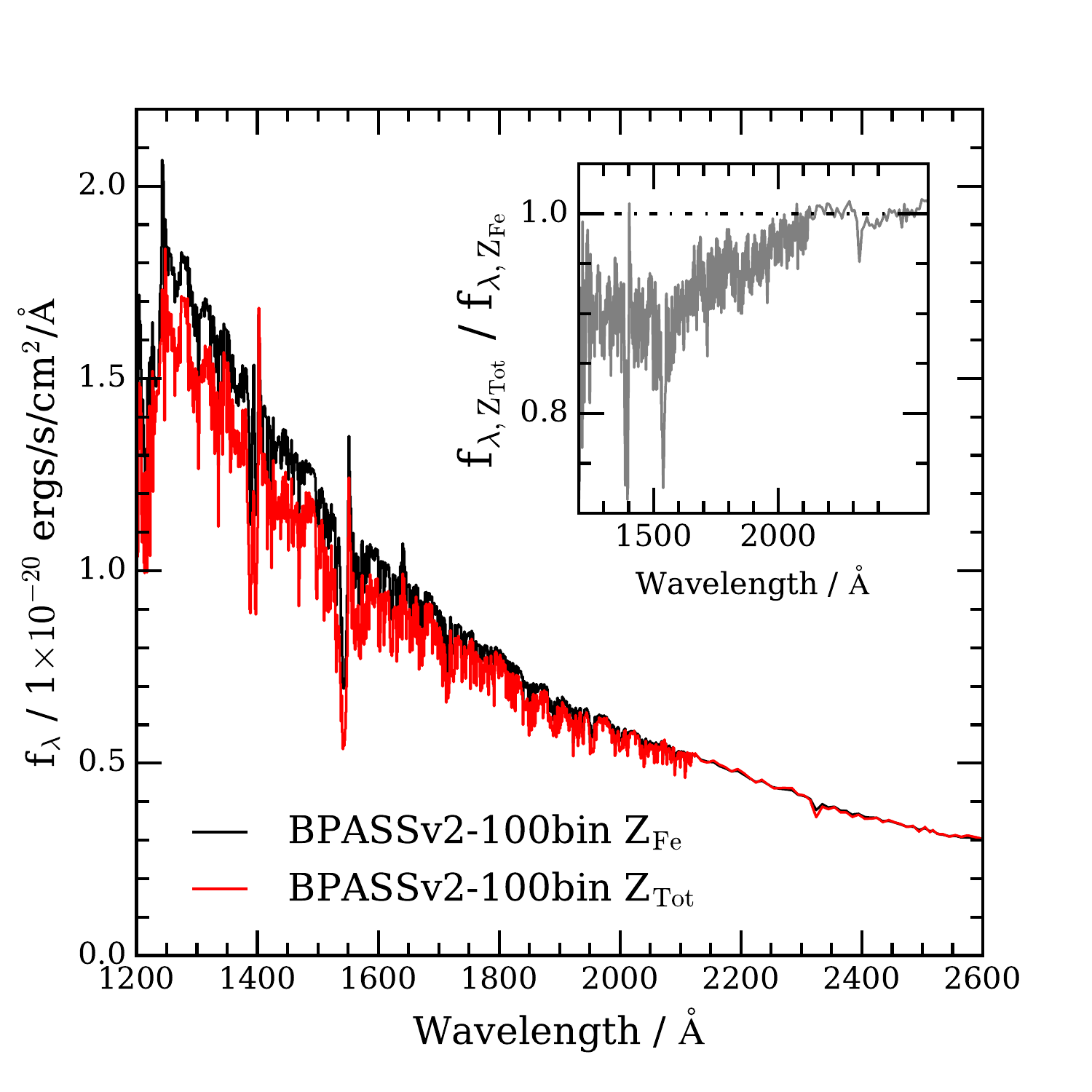}}
        \caption{Comparison of the average composite spectra of N=73 galaxies with log(M/M$_{\odot}$)$\geq9.0$ (for the BPASSv2-100bin model), assuming both $\rm{Z}_{*}=\rm{Z}_{\rm{Fe}}$ (black) and $\rm{Z}_{*}=\rm{Z}_{\rm{Tot}}$ (red).
        Since the $\alpha/$Fe ratio is a increasing function of stellar mass, we chose a subsample of the most massive galaxies to emphasize the effect.
        The inset shows the ratio of the two spectra (f$_{\lambda, \rm{Z_{\rm{Tot}}}}$ / f$_{\lambda, \rm{Z_{\rm{Fe}}}}$). The median ratio of at $\lambda < 2000\rm{\AA}$ is 0.92.}
        \label{fig_zFe_zAlpha_stacked_spectrum}
    \end{figure}

To summarize, as a consequence of the above considerations, we used the mass fraction in Fe, relative to solar, rather than the total mass fraction in metals, to select the appropriate metallicity BPASSv2 model when constructing the galaxy SEDs.
The effect of this choice on the UV SED is illustrated in Fig. \ref{fig_zFe_zAlpha_stacked_spectrum}, which shows the average composite BPASSv2-100bin spectra of all N=73 galaxies with log(M/M$_{\odot}$) $\geq 9.0$, assuming both $\rm{Z}_{*}=\rm{Z}_{\rm{Fe}}$ and $\rm{Z}_{*}=\rm{Z}_{\rm{Tot}}$, and will be discussed further in Section \ref{sec_beta_m1500}.

\subsection{Photoionization models}\label{sec_cloudy}

Since the BPASSv2 models only account for the stellar component of each spectrum, we separately modelled the nebular contribution for each galaxy.
The nebular continuum is a result of free-free, free-bound and two-photon emission from \hii \ regions in the galaxy. 
To model the nebular continuum and emission spectrum we use the photoionization code Cloudy \citep[v13.03;][]{ferland2013}.
We ran Cloudy assuming a plane-parallel geometry and considered the following four key input parameters: (i) the shape of the incident ionizing radiation field, (ii) the metallicity of the nebular gas, (iii) the electron density and (iv) the ionization parameter.
Of these, we directly set (i) and (ii) from the FiBY simulation data.
For the shape of the incident radiation field we input the BPASSv2 stellar continuum of each galaxy, thus the nebular spectrum is modelled by assuming the entire galaxy acts as a single \hii \ region.
Following the discussion in Section \ref{sec_fe_alpha_metal}, we used the UV-weighted alpha-element stellar metallicity to set the nebular abundances, scaling to the \citet{asplund2009} solar abundance set.
This explicitly assumes that the gas phase metallicity traces the metallicity of the most recently formed stars.
To set appropriate values of electron density and ionization parameter at $z=5$ we used recent near-IR spectroscopic observations of star-forming galaxies at $z\simeq2-3$ which have illustrated that the typical values of these parameters are evolving from $z=0$ out to higher redshifts \citep[e.g][]{steidel2014, sanders2016,cullen2016, strom2017}.
For example, \citet{strom2017} have presented a detailed analysis of $\rm{N} \sim 380$ galaxies at $z\sim2.3$ and concluded that the typical value of electron density is $\simeq 300$ cm$^{-3}$ \citep[representing, roughly, an order of magnitude increase over typical SDSS star-forming galaxies; see also][]{sanders2015a}, with ionization parameters in the range $-3.1 < \rm{log}(\rm{U}) < -2.5$, consistent with the best-fitting value of $\rm{log}(\rm{U}) = -2.8$ from \citet{steidel2016} and again representing an increase on typical SDSS values \citep[log(U) $\simeq -3.2$;][]{liang2006, kewley2008}.

We made the assumption that the average $z\sim2.3$ values of these parameters are appropriate to use for our $z=5$ galaxies based on the following reasoning.
Firstly, though the range of masses and SFRs of our sample are not representative of the galaxies from \citet{strom2017} (which have medians of 1.0 x 10$^{10}$ M$_{\odot}$ and 24 M$_{\odot}\rm{yr}^{-1}$, factors $\sim$ 40 and 20 larger respectively) the typical values of sSFR are within a factor 2 (4.8 Gyr$^{-1}$ for our sample compared to 2.4 Gyr$^{-1}$), and one of the key results presented in \citet{strom2017} is that the sSFR is correlated with the degree of excitation and the evolution in optical line ratios they observe.
Unfortunately, no explicit relations between sSFR and either log(U) or $n_{e}$ exist in the literature to our knowledge, and we do not attempt to second guess the potential evolution of these parameters at $z \simeq 5$.
Secondly, the metallicities of our sample (both Fe/H and $\alpha$/H) follow their best-fitting values at $z \simeq 2 - 3$ in the relevant mass range. 
Given the similarities in sSFRs and metallicities, we made the assumption the other physical conditions in \hii \ regions at $z=5$ are also plausibly equivalent.
Therefore, we adopted an electron density of $n_{e} = 300$ cm$^{-3}$ and assumed log(U)$=-2.8$.
We briefly discuss, in Section \ref{sec_discussion}, how our results are affected under the assumption that physical conditions in \hii \ regions continue to evolve towards even more extreme values of these parameters at $z \gtrsim 3$.

\begin{figure}
        \centerline{\includegraphics[width=\columnwidth]{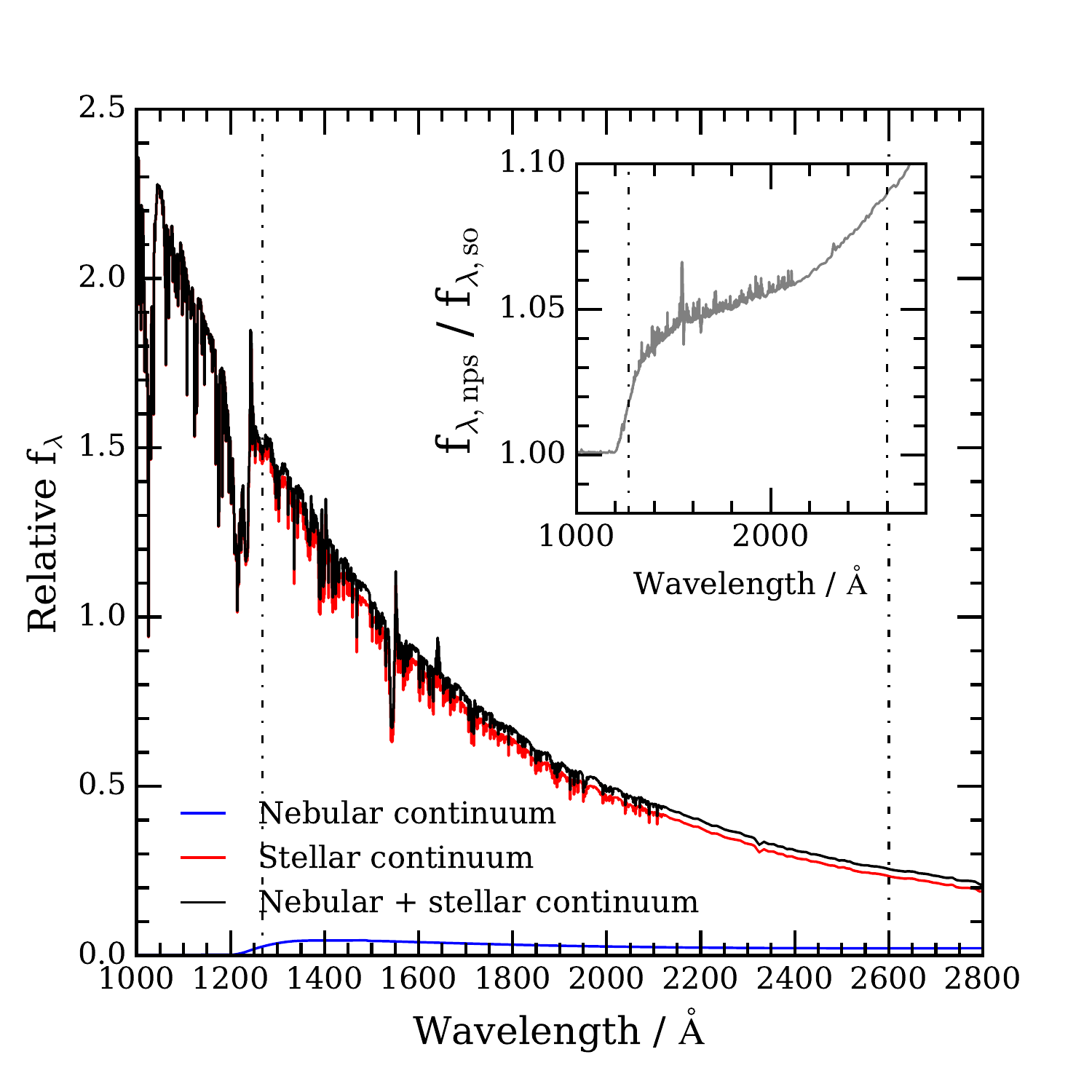}}
        \caption{An example showing the effect of adding nebular continuum emission to the stellar SEDs. This particular example illustrates a typical galaxy from our sample (in terms of mass, SFR, sSFR and metallicity) assuming the BPASSv2-300bin SPS model and $\rm{log}(\rm{U}) = -2.8$ (see text for details). The BPASSv2-300bin stellar continuum is shown in red, with the nebular continuum output from Cloudy shown in blue; the black line is the combined stellar + nebular spectrum. All fluxes are normalized to the $\lambda = 1500 \rm{\AA}$ flux in the stellar-only model. The inset shows the ratio of spectra with and without the nebular contribution. In both axes the vertical dot-dashed lines bracket the wavelength range we used for measuring $\beta$ slopes (see Section \ref{sec_beta_m1500}); in this particular example the median nebular contribution to the final spectrum is $\simeq 6 \%$ in this region.} 
        \label{fig_nebular_continuum}
    \end{figure}

The effect of adding the nebular continuum to the UV spectrum is illustrated in Fig. \ref{fig_nebular_continuum} for a typical galaxy in our sample (M$_{*}$ = 2.8 x 10$^8$ M$_{\odot}$; SFR = 1.3 M$_{\odot}\rm{yr}^{-1}$; sSFR = 4.6 Gyr$^{-1}$; Z$_{\rm{Fe}}$/Z$_{\odot}$ = 0.06).
In this particular example we show a BPASSv2-300bin model SED; the median nebular contribution over the wavelength range within which we measured $\beta$ slopes ($1268 < \lambda < 1950 \rm{\AA}$) is $\simeq 6 \%$, and the overall effect of the nebular continuum is to make the spectrum redder (i.e. increase the value of $\beta$).
We note that although nebular emission is included in our photoionization modelling, nebular emission lines have no effect on our method of constraining the form of the attenuation law, therefore in Fig. \ref{fig_nebular_continuum}  we do not show emission lines, since only the continuum is relevant in this context. 
A detailed discussion of the effect of the nebular continuum on $\beta$ across the whole sample is given in Section \ref{sec_beta_m1500}.

To summarize, we have modelled the nebular continuum of all galaxies in our sample by using the BPASSv2 stellar SEDs to define the shape of the incident spectrum, assumed the nebular metallicity to be equal to the UV-weighted total stellar metallicity of the galaxy, and used values for electron density and ionization parameter typical of star-forming galaxies at $z\simeq 2-3$.
Therefore, for each galaxy, we have both the intrinsic stellar spectrum, and the combined stellar $+$ nebular spectrum.
As we will discuss later in the paper, the pure stellar spectrum provides a useful limit for our study of plausible attenuation curves, since it represents the `maximally blue' SEDs of galaxies at $z=5$ from the FiBY simulations, based purely on their star-formation histories and metallicities.

\section{Intrinsic $\bmath{\beta}$ and M$_{\rm{1500}}$ Distributions}\label{sec_beta_m1500}

As will be discussed in detail in Sections \ref{sec_dust_model} and \ref{sec_fitting}, in order to constrain the form of the dust attenuation curve, we have compared our simulations to two observed properties of $z=5$ galaxies: (i) the UV continuum slope ($\beta$) and (ii) the volume density of galaxies as a function of their absolute magnitude at $\lambda = 1500 \rm{\AA}$ (M$_{\rm{1500}}$). 
In this section we describe the intrinsic values we measured for these parameters from the synthetic SEDs, as well as discussing how the various different assumptions we have made (e.g. stellar metallicity, nebular modelling) systematically affect these observables.

\subsection{UV continuum slope ($\beta$)}

The UV continuum slope of a galaxy is defined as a power-law expressed as
\begin{equation}
f_{\lambda} \propto \lambda^{\beta}
\end{equation}
where $\beta$ is the power-law index \citep[e.g.][]{bouwens2010,dunlop2012,dunlop2013}.
In the absence of dust, the intrinsic UV continuum is sensitive to the age and metallicity of the stellar population, with older and more metal rich populations having redder continuum slopes.

\subsubsection{Intrinsic stellar UV continuum}\label{sec_stellar_UV_cont}

\begin{figure}
        \centerline{\includegraphics[width=\columnwidth]{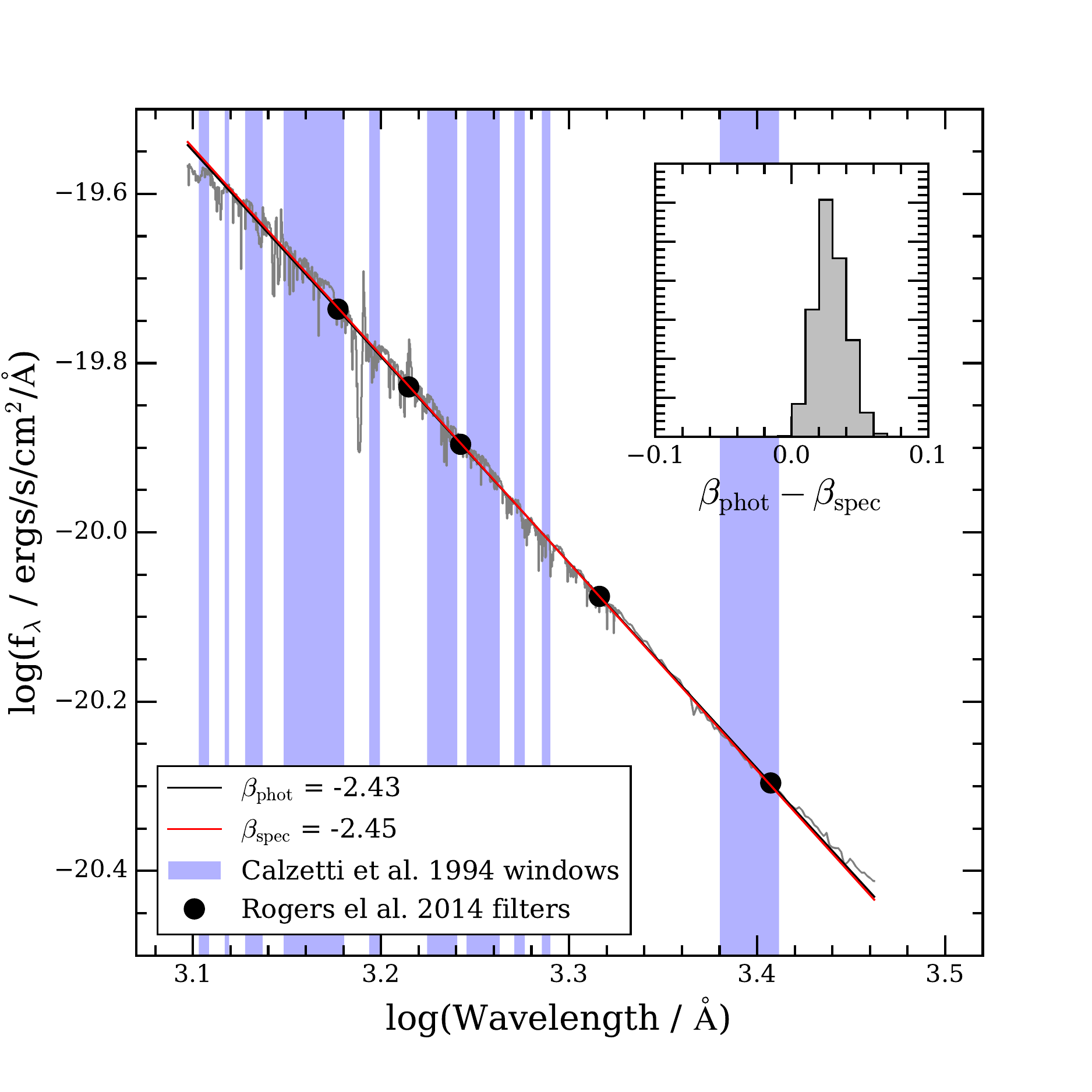}}
        \caption{An example comparing the method of measuring $\beta$ by fitting the UV spectrum directly, to the photometric fitting method used by \citet{rogers2014} to observationally constrain $\beta$ at $z=5$. The same galaxy spectrum is shown in grey and the black points are the synthetic photometry in the \citet{rogers2014} filters. The black line shows the fit to the photometric data, and the red line shows the fit to the spectrum using the \citet{calzetti1994} windows (shown as shaded blue regions). The inset shows the distribution of $\beta_{\rm{phot}} - \beta_{\rm{spec}}$ for the N=498 galaxies across all four BPASSv2 models; the median offset is $\beta_{\rm{phot}} - \beta_{\rm{spec}} = 0.03$.}
        \label{fig_measure_beta}
    \end{figure}

To measure $\beta$ for our synthetic SEDs we performed a linear fit to log$f_{\lambda}$ versus log$\lambda$ over the wavelength range $1268 < \lambda < 2580$, masking out stellar and interstellar absorption features using the windows defined in \citet{calzetti1994}, we refer to this as a `spectroscopic' $\beta$ measurement.
As we will describe in Section \ref{sec_fitting}, we used $\beta$ slopes of $z=5$ galaxies reported in \cite{rogers2014} as our observational comparison.
These authors measured $\beta$ by fitting a power-law to five photometric bands which sampled the rest-frame UV spectrum in the wavelength range $\lambda \approx 1500 - 2600 \rm{\AA}$. 
To ensure consistency, we checked that our spectroscopic $\beta$ measurements were consistent with the \citet{rogers2014} photometric method.
We found that both methods are consistent for idealized data (see Fig. \ref{fig_measure_beta}), with the spectroscopic $\beta$ measurements systematically smaller by a median of $\Delta \beta = 0.03$ across the whole sample.
For our sample, we chose to use the spectroscopically-measured $\beta$ values rather than explicitly mimic observational methods.

\begin{figure}
        \centerline{\includegraphics[width=\columnwidth]{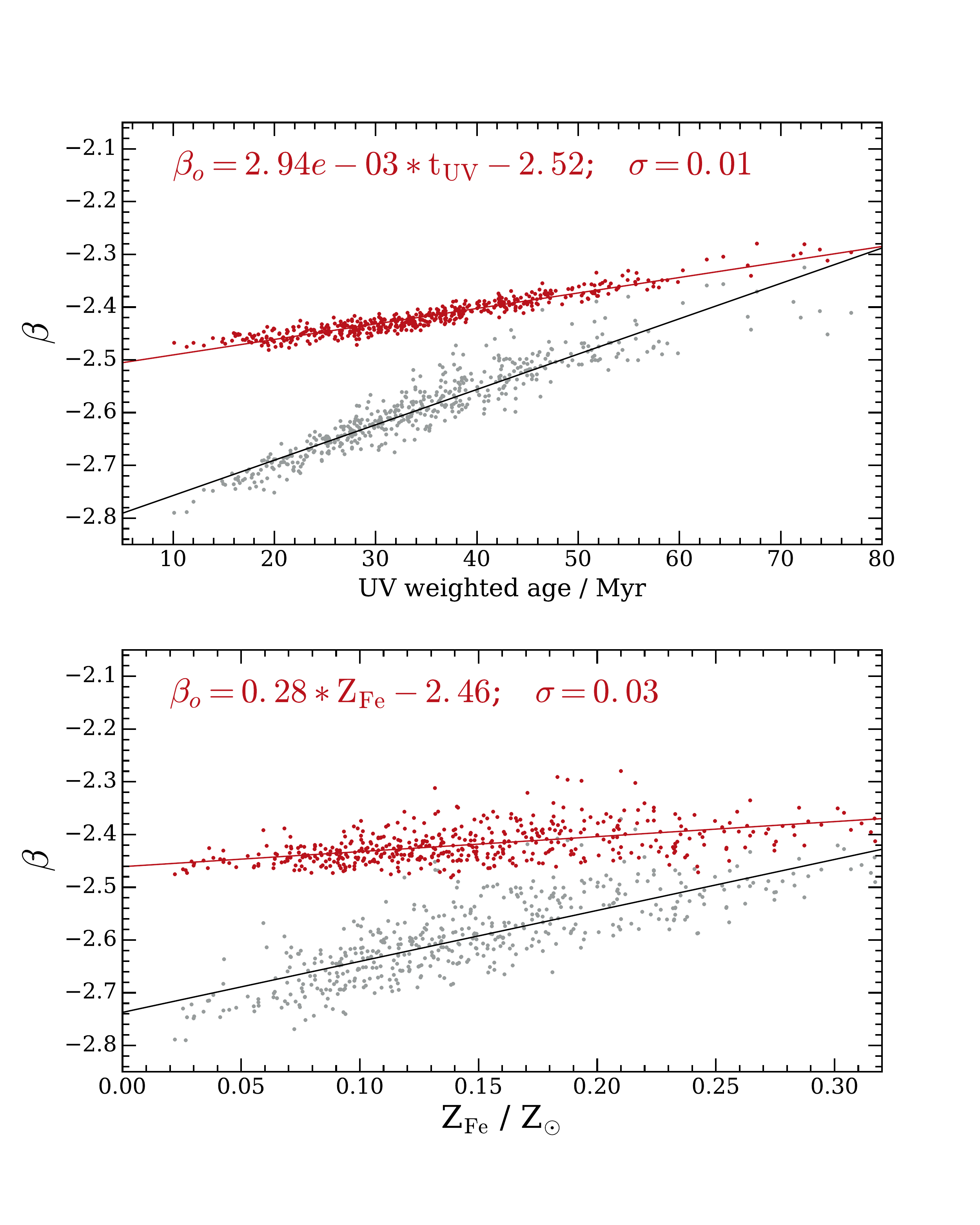}}
        \caption[]{Illustration of the effect of UV-weighted age (upper panel) and UV-weighted Fe-based metallicity (lower panel) on the intrinsic stellar UV continuum slopes for SEDs generated with the BPASSv2-100bin models. 
        In each panel the intrinsic stellar UV continuum slopes ($\beta_{i}$) are shown as grey points and the slopes modified by nebular continuum emission ($\beta$, see Section \ref{sec_neb_beta_effect}) are shown as red points.
        In each panel the best-fitting linear relationship is shown as the heavy solid line with the equation and dispersion about the relationship for the stellar+nebular case given in each panel.}
        \label{fig_beta_uvage_zfe}
    \end{figure}

Fig. \ref{fig_beta_uvage_zfe} shows how the intrinsic stellar UV continuum ($\beta_{i}$) of our sample varies as a function of the UV-weighted age (defined as age weighted by the luminosity at $1500\rm{\AA}$ of each star particle) and the UV-weighted Fe-based metallicity (Z$_{\rm{Fe}}$).
We note that the data in Fig. \ref{fig_beta_uvage_zfe} are taken from the SEDs generated with the BPASSv2-100bin models, however, although the normalization of $\beta_{i}$ differs slightly, the trends are the same across all SPS models.
As expected, the value of $\beta_{i}$ increases with increasing UV-weighted age and UV-weighted metallicity. 
Moreover, we find that, at $z=5$, $\beta_{i}$ is more strongly correlated with the UV-weighted age within the range $10 - 80$ Myr probed by our sample, with the scatter in $\beta_{i}$, at fixed UV-weighted age, $\approx 1/3 \times$ the scatter at fixed metallicity.
$\beta_{i}$ also correlates positively with mass-weighted age, although with a larger dispersion due to the mass-weighted age probing older stellar populations which contribute less to the UV SED.
For the BPASSv2-100bin model SEDs shown in Fig. \ref{fig_beta_uvage_zfe}, the median $\beta_{i}$ across all galaxies is $-2.60$, for the other SPS models the medians are $-2.55$, $-2.63$ and $-2.58$ for BPASSv2-100, BPASSv2-300bin and BPASSv2-300 respectively (see Table \ref{tab_beta_values}). 
Independent of SPS model, the dispersion of $\beta_{i}$ values is $\sigma_{\beta_{i}} \approx 0.08$. 

Given the correlation between metallicity and $\beta_{i}$ shown in Fig. \ref{fig_beta_uvage_zfe}, it is clear that adopting Fe-based rather than total metallicities for generating stellar SEDs can have a strong impact on the shape of the UV continuum (see also Fig. \ref{fig_zFe_zAlpha_stacked_spectrum}).
We find that the offset between $\beta_{i, \rm{Fe}}$ and $\beta_{i, \rm{Tot}}$ is an increasing function of stellar mass, consistent with the trend shown in the right-hand panel of Fig. \ref{fig_zFe_zAlpha}.
For the BPASSv2-100bin model the offset can be as large as $\delta \beta_{i} \approx 0.3$ at the highest masses, with a median of 0.13; the median offsets across all other models are 0.02, 0.23 and 0.09 for BPASSv2-100, BPASSv2-300bin and BPASSv2-300 respectively.
This emphasizes the importance, in future work (including SED fitting to observations) of adopting the correct stellar metallicity when fitting the UV continuum, which, as we have previously discussed, is not equal to the total mass-weighted metallicity in the case of variable $\alpha$/Fe ratios.

    \begin{figure}
        \centerline{\includegraphics[width=\columnwidth]{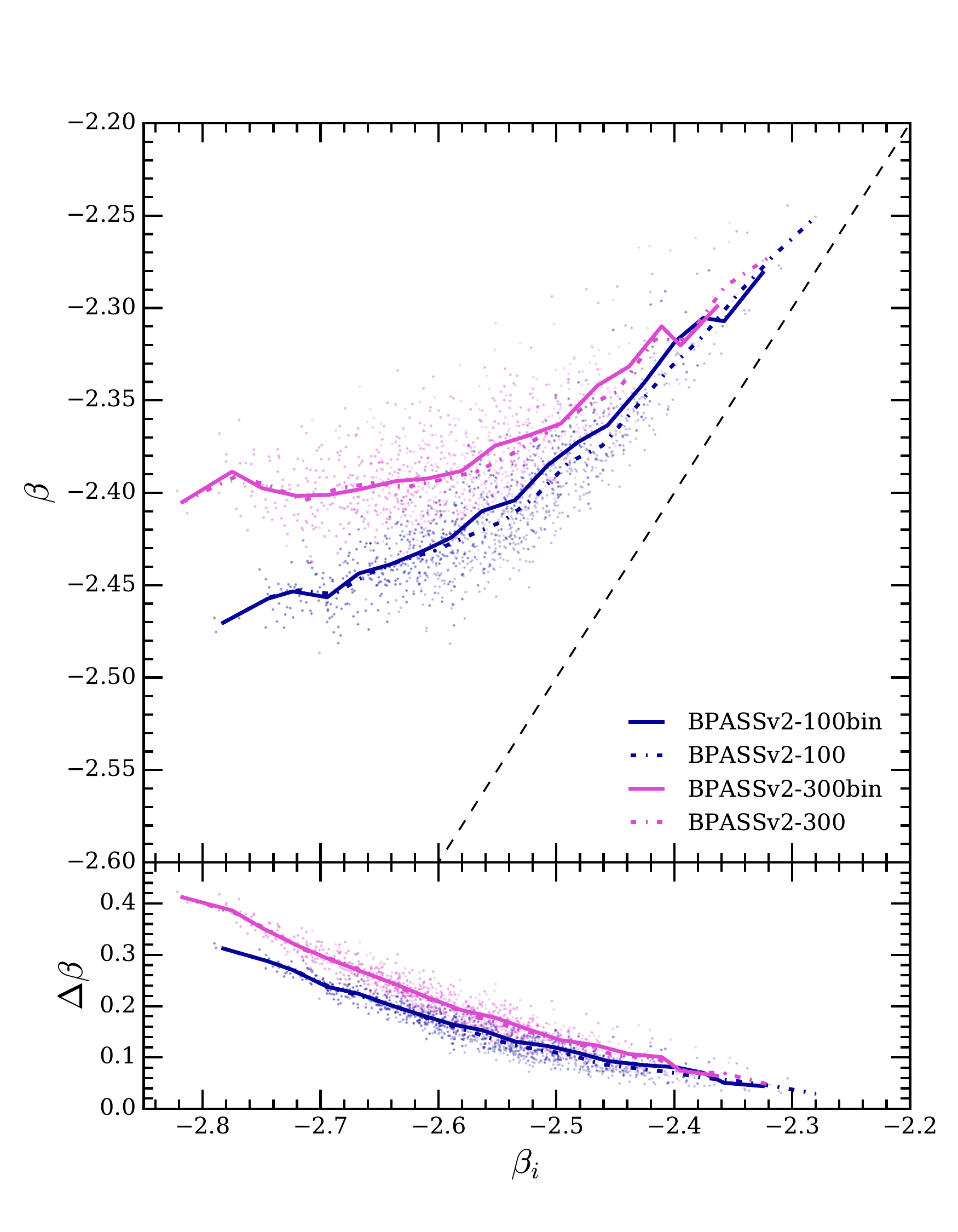}}
        \caption[]{The effect of nebular continuum emission on the intrinsic stellar UV continuum slopes ($\beta_{\rm{i}}$) of the galaxy sample across all SPS models.
        The top panel shows $\beta_{i}$ versus $\beta$ with the black dashed line indicating a 1:1 relationship.
        The bottom panels shows $\Delta \beta$ ($= \beta - \beta_{i}$) versus $\beta_{i}$.
        In both panels the solid and dot-dashed lines show the running medians for each of the four SPS models as indicated in the legend, with the scatter points representing the individual galaxies.} 
        \label{fig_nebular_beta_effect}
    \end{figure}

    \begin{figure} 
        \centerline{\includegraphics[width=\columnwidth]{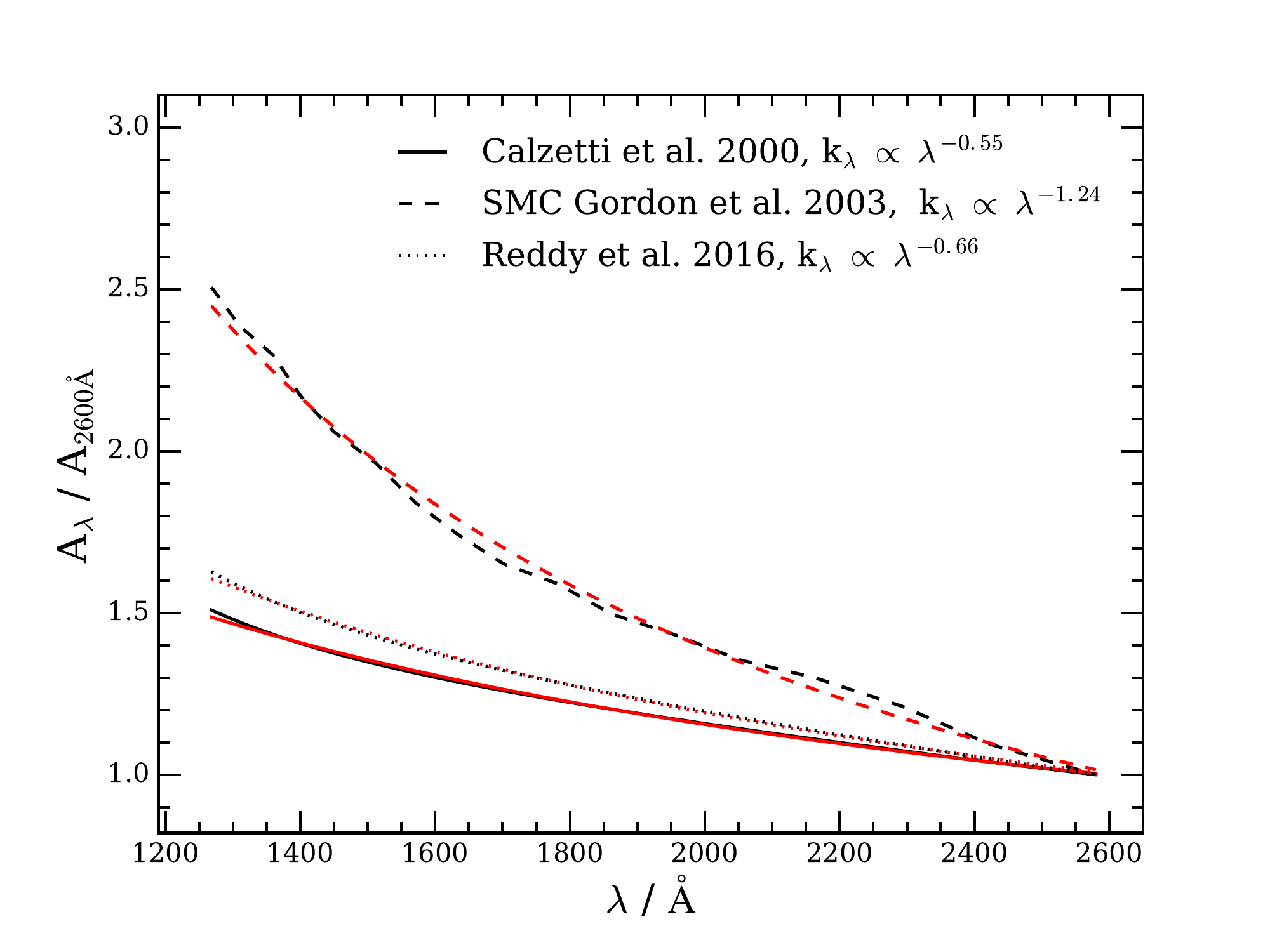}}
        \caption{Illustration of three simple parameterizations of the shape of the attenuation/extinction curve in the UV compared to three observationally-derived curves from the literature. The black curves show the observations, and over plotted in red are a power-law fits of the form $\lambda^n$. The `greyest', and most commonly adopted, attenuation curve is \citet{calzetti2000} which is roughly equivalent to a power-law with A$_{\lambda} \propto \lambda^{-0.55}$. The attenuation curve of \citet{reddy2015} and the extinction curve of \citet{gordon2003} (SMC) are steeper, with power-law exponents of $-0.66$ and $-1.24$ respectively. The `steepness' of these curves significantly affects how the shape of the UV continuum changes for an equivalent amount of absolute reddening (see Section \ref{sec_fitting}).} 
        \label{fig_klam}
    \end{figure}{}

\subsubsection{Effect of the nebular continuum}\label{sec_neb_beta_effect}

As illustrated in Figs. \ref{fig_nebular_continuum} and \ref{fig_beta_uvage_zfe}, the nebular continuum has a non-negligible effect on the UV continuum shape of individual galaxies.
Here we discuss the effect of the nebular continuum on $\beta_{i}$ across our full sample.
For the remainder of this paper, $\beta$ refers to the UV continuum slope of the galaxies including the nebular continuum.

Fig. \ref{fig_nebular_beta_effect} shows the relationship between $\beta$ and $\beta_{i}$ across all four BPASSv2 SPS models.
The addition of the nebular continuum acts to redden the intrinsic stellar slopes (i.e. leads to larger values of measured $\beta$).
The effect is strongest for the galaxies with the steepest (or bluest) stellar UV continuum slopes, or, equivalently, the galaxies with the lowest metallicities and youngest UV-weighted ages.
This is unsurprising since young, low metallicity stellar populations generate harder ionizing spectra and hence a stronger nebular continuum.
For the bluest stellar continua the effect can be as large as $\Delta \beta = 0.4$, but decreases to $\Delta \beta \approx 0.05$ at $\beta_{i} \approx -2.35$, with a medians of $\Delta \beta \approx 0.15 - 0.25$ across all models.
In other words, the effect of the nebular continuum on $\beta$ is of the order $\approx 2 - 15 \%$, with an average of $\approx 10 \%$.

It is also evident from Fig. \ref{fig_nebular_beta_effect} that, while the difference between binary and single-star models is minor, $\Delta \beta$ is systematically larger for the BPASSv2-300 models. 
At its maximum, the magnitude of this effect represents a difference in $\Delta \beta$, at fixed $\beta_{i}$, of $\approx 0.05$; interestingly, this is large enough to mean that while the median $\beta_{i}$ is bluer for the BPASSv2-300 compared to the BPASSv2-100 models, the situation is reversed for the $\beta$ medians (see Table \ref{tab_beta_values}).

Finally, across our sample, including the nebular continuum has the effect of reducing the scatter in intrinsic $\beta$ values by a factor $\approx 2$ (see Table \ref{tab_beta_values}) and significantly flattens the relations between $\beta$ and metallicity and UV age (Fig. \ref{fig_beta_uvage_zfe}).
The median $\beta$ across all SPS models is $\approx -2.40$, and we find that the minimum value of $\beta$ in the case of no dust attenuation at $z=5$ is $\beta_{\rm{min}} = -2.50$.
This theoretical dust-free $\beta$ limit is consistent with the work of \citet{dayal2012} and \citet{wilkins2013} at $z = 6 - 8$ and, as we will return to later in the paper, significantly different to the value of $\beta_{\rm{min}} = -2.23$ assumed for the \citet{meurer1999} IRX $-\beta$ relation.

\subsection{UV continuum magnitude}

The UV continuum magnitude (M$_{1500,i}$) is primarily sensitive to the star-formation history of a galaxy over the last 10 - 100 Myr \citep{kennicutt2012}.
As mentioned in Section \ref{sec_fiby_data}, we calculated the intrinsic absolute UV magnitude magnitude (M$_{1500,i}$) by integrating the rest-frame SEDs over a top-hat filter of width $100 \rm{\AA}$ at a central wavelength of $1500 \rm{\AA}$.

Since our sample was defined by requiring an intrinsic stellar magnitude (i.e. dust free, no nebular contribution) M$_{1500,i} \leq -18.0$, the faintest galaxies across all SPS models have roughly this magnitude (see Section \ref{sec_fiby_data}); the (median, maximum) M$_{1500,i}$ across all models are $(-18.88, -23.25)$, $(-18.77, -23.16)$, $(-18.99, -23.36)$ and $(-18.87, -23.26)$ for the BPASSv2-100bin, BPASSv2-100, BPASSv2-300bin and BPASSv2-300 models respectively.
In the $\approx 3.7 \times 10^4$ Mpc$^3$ volume, we do not find any galaxies intrinsically brighter than M$_{1500,i} \approx -23.4$.

As with the $\beta$ values, the choice of stellar metallicity also affects the intrinsic stellar M$_{1500,i}$ distributions, in the sense that a galaxy with a low SPS model metallicity will be brighter in the UV, due to reduced photospheric line blanketing.
Again the effect is strongest for the binary models and models with a higher mass IMF cutoff, but not strong enough to affect the results presented in this paper; for the BPASSv2-300 models, the median offset in M$_{1500,i}$ is 0.23 mag.
Finally, the effect of the nebular continuum on M$_{1500,i}$ is also small; across all models, adding nebular continuum increases the brightness of the SEDs by $\lesssim 0.1$ mag.

    \begin{table}
        \caption{The medians and standard deviations of the intrinsic UV stellar continuum slope ($\beta_{i}$), the stellar + nebular UV continuum slope ($\beta$) and $\Delta \beta$ ($= \beta - \beta_{i}$) for all four SPS models (all in the case of zero dust). $\Delta \beta$ is not the same as the difference in medians because the $\beta$ distributions are skewed.}\label{tab_beta_values}
        \begin{tabular}{lccc}
            \hline
            SPS Model & $\beta_{i}$ & $\beta$ & $\Delta \beta$ \\
            \hline
            BPASSv2-100bin & $-2.60 \pm 0.08$ & $-2.42 \pm 0.04$ & 0.17 $\pm$ 0.05\\
            BPASSv2-100 & $-2.55 \pm 0.09$ & $-2.42 \pm 0.04$ & 0.14 $\pm$ 0.06\\
            BPASSv2-300bin & $-2.63 \pm 0.08$ & $-2.39 \pm 0.03$ & 0.23 $\pm$ 0.07\\
            BPASSv2-300 & $-2.58 \pm 0.08$ & $-2.39 \pm 0.03$ & 0.19 $\pm$ 0.07\\
            \hline
        \end{tabular}
    \end{table}

\section{Modelling the dust attenuation}\label{sec_dust_model}

Having defined the intrinsic properties of simulated galaxies at $z=5$, we now turn to exploring models for dust attenuation, which should map these intrinsic properties to their observed values.
In particular, we are interested in whether deviations from a Calzetti-like attenuation curve are required to achieve this match, since evidence (both observed and simulated) for a steeper attenuation curve, more like the SMC extinction curve, at high redshifts has been reported in the recent literature \citep[e.g.][see further discussion below]{capak2015,bouwens2016,mancini2016}.
To do this, we considered two simple models for the dust attenuation: (i) a total birth-cloud extinction model (TEx) and (ii) a \citet{charlot2000}-like model (CF2000) which we describe in detail below.

    \begin{figure*}
        \centerline{\includegraphics[width=8in]{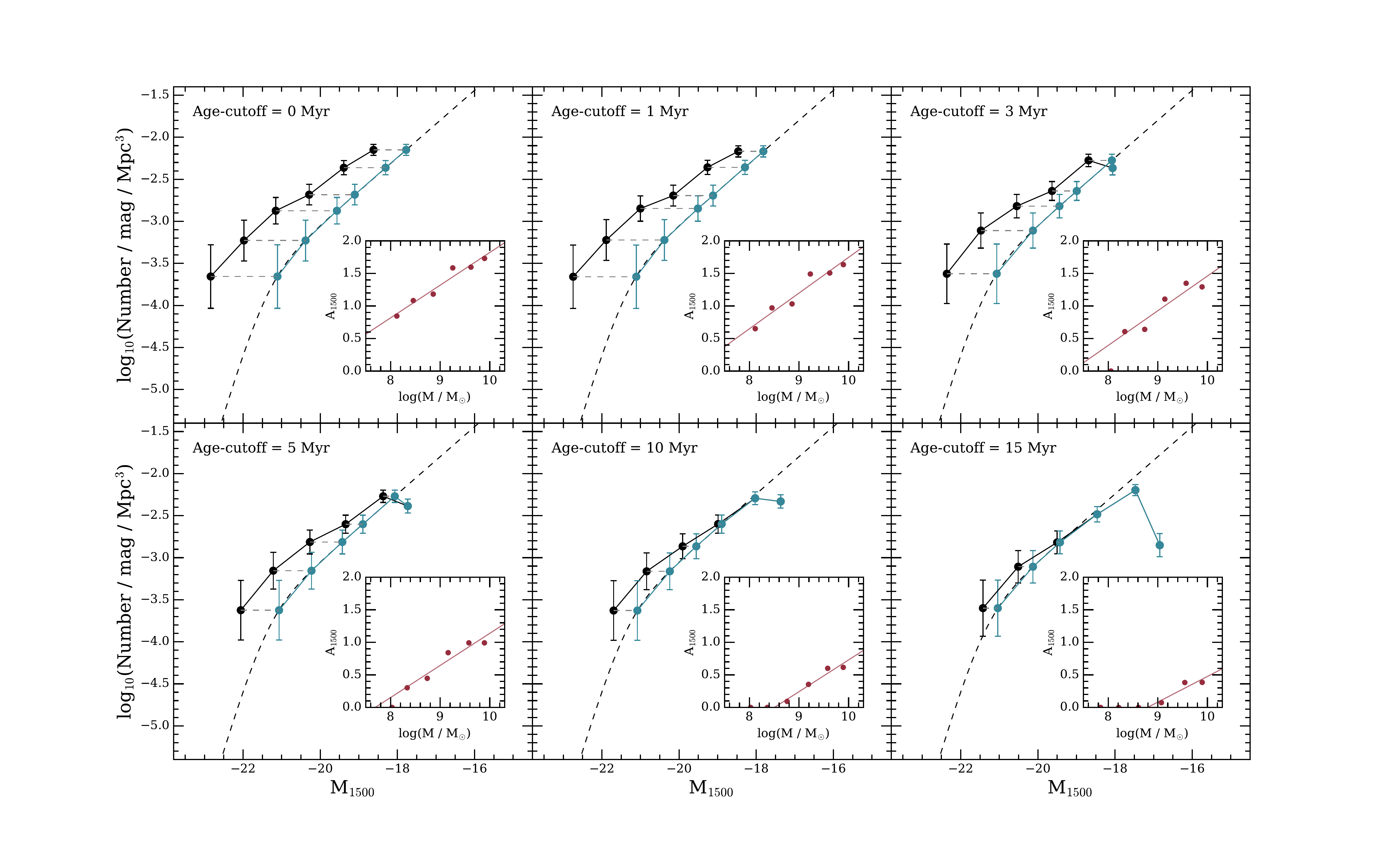}}
        \caption[]{Illustrating the method for deriving the A$_{1500}$ versus log(M/M$_{\odot}$) relation for the TEx model using the $z=5$ UV luminosity function of \citet{bowler2015} (dashed black line).
        This illustrative example uses the BPASSv2-100bin models. 
        Each panel shows one of the six birth-cloud ages described in Sec. \ref{sec_tbex_model} as indicated in the upper left-hand corner. 
        In each panel, the black connected points show the intrinsic luminosity function in six bins and the turquoise points show the original points after applying the necessary A$_{1500}$ correction to move them onto the observed relation.
        In cases where the intrinsic luminosity function of our model falls below the observed relation we assign A$_{1500}=0$.
        The errors on the points are the Poisson error in each bin.
        The inset in each panel shows the implied A$_{1500}$ - log(M/M$_{\odot}$) relation after making this correction, with the red solid line a linear fit through the points with A$_{1500}>0$.}
        \label{fig_bc_model_lf}
    \end{figure*}

\begin{table*}
    \caption{The best-fitting values of the reddening law slope ($n$) for the TEx model, including both the stellar + nebular continuum and stellar continuum only cases.}\label{tab_bc_model_chi2}
    \begin{tabular}{l c c c c c}
        \hline
        SPS Model & Birth Cloud Age / Myr & $n_{\rm{neb}}$ & $\chi^2_{\rm{neb}}$ & $n_{\rm{stellar}}$ & $\chi^2_{\rm{stellar}}$ \\  
        \hline
        \hline
        BPASSv2-100bin & 0 &  $-0.45$ & 0.33 & $-0.60$ & 1.94\\ 
        BPASSv2-100bin & 1 &  $-0.47$ & 0.19 & $-0.60$ & 3.95\\ 
        BPASSv2-100bin & 3 &  $-0.51$ & 0.49 & $-0.62$ & 3.24\\ 
        BPASSv2-100bin & 5 &  $-0.60$ & 1.83 & $-0.69$ & 4.58\\ 
        BPASSv2-100bin & 10 &  $-0.73$ & 12.84 & $-0.86$ & 10.47\\ 
        BPASSv2-100bin & 15 &  $-0.53$ & 20.99 & $-0.87$ & 17.57\\ 
        \hline
        BPASSv2-100 & 0 &  $-0.48$ & 0.12 & $-0.58$ & 1.95\\ 
        BPASSv2-100 & 1 &  $-0.52$ & 0.74 & $-0.59$ & 3.80\\ 
        BPASSv2-100 & 3 &  $-0.55$ & 1.54 & $-0.60$ & 3.17\\ 
        BPASSv2-100 & 5 &  $-0.67$ & 5.75 & $-0.67$ & 4.09\\ 
        BPASSv2-100 & 10 &  $-0.63$ & 13.66 & $-0.75$ & 8.69\\ 
        BPASSv2-100 & 15 &  $-0.33$ & 17.34 & $-0.70$ & 15.00\\ 
        \hline
        BPASSv2-300bin & 0 &  $-0.39$ & 1.04 & $-0.58$ & 2.03\\  
        BPASSv2-300bin & 1 &  $-0.43$ & 0.23 & $-0.58$ & 3.37\\ 
        BPASSv2-300bin & 3 &  $-0.50$ & 0.47 & $-0.64$ & 3.65\\ 
        BPASSv2-300bin & 5 &  $-0.64$ & 2.15 & $-0.72$ & 5.10\\ 
        BPASSv2-300bin & 10 &  $-0.74$ & 17.17 & $-0.93$ & 12.45\\ 
        BPASSv2-300bin & 15 &  $-0.45$ & 25.39 & $-0.90$ & 21.55\\ 
        \hline
        BPASSv2-300 & 0 &  $-0.42$ & 0.42 & $-0.56$ & 2.25\\ 
        BPASSv2-300 & 1 &  $-0.47$ & 0.29 & $-0.57$ & 3.38\\ 
        BPASSv2-300 & 3 &  $-0.56$ & 1.87 & $-0.63$ & 3.64\\ 
        BPASSv2-300 & 5 &  $-0.70$ & 7.49 & $-0.71$ & 4.95\\ 
        BPASSv2-300 & 10 &  $-0.61$ & 17.59 & $-0.80$ & 10.90\\ 
        BPASSv2-300 & 15 &  $-0.25$ & 19.72 & $-0.70$ & 18.94\\ 
        \hline
    \end{tabular}
\end{table*}

\subsection{TEx}\label{sec_tbex_model}

The total birth-cloud extinction model (TEx) assumes that, below a critical age, star particles are completely embedded in their birth clouds and suffer total extinction, while at larger ages either their birth clouds have been photo-evaporated by OB star associations, or the OB associations have drifted away or been blown-out of the birth-cloud, and hence the star particles suffer only the global ambient ISM attenuation of the galaxy.

Local estimates of the timescale for molecular cloud destruction are of the order $\sim$ 10 - 30 Myr \citep[][]{blitz1980,blitz2007,mckee2007}, values which are in good agreement with current predictions from hydrodynamical simulations \citep[e.g.][]{colin2013, kortgen2016}; while estimates of the timescales for the ejection of OB associations from their birth clouds are of the order 1 - 3 Myr \citep[e.g.][]{israel1978,leisawitz1988}, consistent with the youngest ages of galactic OB associations \citep[e.g.][]{massey1995}.

Given these observations, we chose to model the following six birth cloud age thresholds $t_{BC} = $ 0, 1, 3, 5, 10 and 15 Myr.
The upper limit of 15 Myr is chosen to be representative of the timescale for molecular cloud dispersion, and to mirror the work of \citet{jimenez2000} who find, using a similar model, that applying this age threshold successfully reproduces the shape of the UV spectrum of the $z=2.73$ Lyman-break galaxy 1512-cB58 \citep{pettini2000}.
The reddening (in magnitudes) as a function of wavelength in this model can be written as,

\begin{equation}\label{equation_tex}
 A(\lambda) =
  \begin{cases}
    \infty       & \quad \text{if } t \leq t_{BC},\\
    \phi \lambda^n  & \quad \text{if } t > t_{BC},\\
  \end{cases}
\end{equation}
where $\phi$ is a linear function in log(M/M$_{\odot}$) given by:

\begin{equation}\label{eq_mass_dep_alam}
\phi = a_1\rm{log}_{10}(M/M_{\odot}) + a_0.
\end{equation}

In this prescription, the overall normalization (or `amount' of dust) is a linearly increasing function of the logarithm of stellar mass.
It is well established in the local Universe that effective dust optical depth increases with stellar mass \citep[e.g.][]{garn2010}; we adopt a linear model, both for simplicity, and because there is no strong evidence at high redshift to favour a more complex functional form.
We modelled the wavelength dependence of the attenuation as a power-law with exponent $n$; this simple parameterization is a good approximation to many observationally derived attenuation curves which don't include a $2175 \rm{\AA}$ `bump' as illustrated in Fig. \ref{fig_klam}.
It is worth mentioning that the results of this paper are not dependent of the presence or size of the $2175 \rm{\AA}$ `bump' feature, since the \citet{calzetti1994} windows are selected to mask out the $2175 \rm{\AA}$ region, and \citet{rogers2014} have demonstrated that their observationally derived $\beta$ values are also insensitive to this feature.

\subsection{CF2000}\label{sec_cf2000_model}

The CF2000 model follows the prescription of \citet{charlot2000}, with the addition of a mass-dependent normalization ($\phi$) as above.
In this model the wavelength-dependent optical depth experienced by a star particle of age (t) is given by,
\begin{equation}\label{eq_cf2000}
 A(t, \lambda) =
  \begin{cases}
    \phi \lambda^n  & \quad \text{if } t \leq t_{BC},\\
    \mu \phi \lambda^n  & \quad \text{if } t > t_{BC},\\
  \end{cases}
\end{equation}
where $\mu$ defines the fraction of the total dust optical depth contributed by the ambient ISM, $n$ is the exponent of the reddening curve, $\phi$ is the overall normalization and $t_{BC}$ is the age of transition between birth cloud and ambient ISM optical depth.
Again we parameterized the normalization (or `amount' of dust) $\phi$ as a linear function of the logarithm of stellar mass as in Eq. \ref{eq_mass_dep_alam}.

The difference here is that, instead of suffering total extinction, stars within their birth cloud suffer additional attenuation with respect to stars in the ambient ISM of a galaxy.
Using a sample of local starburst galaxies, \citet{charlot2000} estimate $\mu \approx 1/3$, $n \approx -0.7$ and $t_{BC} \approx 10$ Myr, however, as will be described in more detail below, we kept all three parameters free when attempting to fit our simulation data to the observed LF and CMR.

    \begin{figure*}
        \centerline{\includegraphics[width=8.5in]{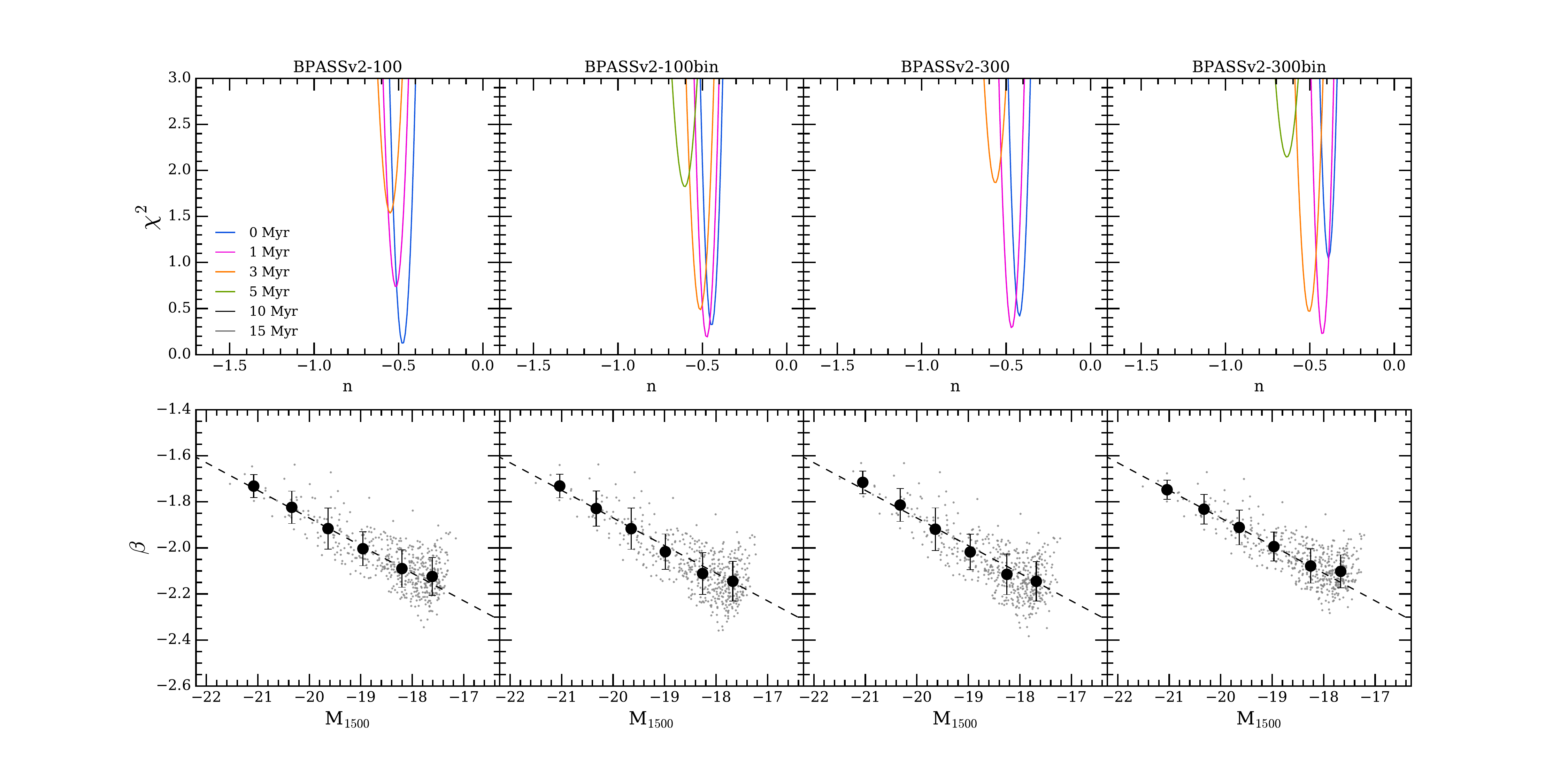}}
        \caption[]{The top panels show $\chi^2$ as a function of the slope of the dust attenuation law ($n$) and the bottom panels show the CMR at the value of the best-fitting solution. Each column represents one of the four BPASSv2 SPS models.
        The top panels show $\chi^2$ versus $n$ for each of the six values of age cutoff adopted in the TEx model. 
        Depending on the model, only a subset are visible as the minimum $\chi^2$ values of the other models are too large to fit within the axis range.
        In the bottom panels, the small grey points show the individual galaxies in our sample and the large black circles shown the medians in six bins. 
        The errors on the points represent the standard deviation of $\beta$ in each bin.
        The dashed black line is the observed $z=5$ CMR from \cite{rogers2014}.}
        \label{fig_bc_model_cmr}
    \end{figure*}

\section{Fitting to the LF and CMR}\label{sec_fitting}

In this section we present the results of using these two dust models to fit to the observed UV galaxy luminosity function and CMR at $z=5$.
The focus of this section is to constrain the dust model parameters (specifically the reddening slope $n$) before, in Section \ref{sec_irx_relation}, we discuss how the predictions of our best-fitting dust models compare to the observed IR properties of high-redshift galaxies.

\subsection{TEx}

For the TEx model, the mass normalization $\phi$ was derived by directly fitting to the $z=5$ luminosity function of \citet{bowler2015}.
The method is illustrated in Fig. \ref{fig_bc_model_lf}: for each birth-cloud age and SPS model, the sample was split into six bins of M$_{1500}$, the number density of galaxies in each bin was determined, and the value of A$_{1500}$ required to match the luminosity function calculated.
A value of A$_{1500}$ = 0 was assigned to bins in which the number density fell below the observed relation. 

We were then able to fit the relationship between A$_{1500}$ and the median stellar mass in each bin.
The inset panels of Fig. \ref{fig_bc_model_lf} show linear fits to A$_{1500}$ versus stellar mass (bins with A$_{1500}$ = 0 were excluded from the fitting).
Using this relation we can then rewrite Equation \ref{equation_tex} for $t > t_{\rm{BC}}$ as,
\begin{equation}\label{eq_alam_a1500}
\rm{A}(\lambda, M) = \rm{A}_{1500}(M) \left(\frac{\lambda}{1500}\right)^n,
\end{equation}
where M $\equiv$ log(M/M$_{\odot}$), leaving the slope of the attenuation law ($n$) as the only model parameter to be fitted.

To constrain the slope of the attenuation law, the CMR of our sample was compared to the observed CMR of \citet{rogers2014} by varying the value of $n$, applying Equation \ref{eq_alam_a1500}, and re-calculating $\beta$ and M$_{1500}$.
For a giving mass normalization, the change in $\beta$ can be expressed as,
\begin{equation}\label{eq_beta_vs_A_calz00}
\Delta \beta = \beta_{\rm{obs}} - \beta_{\rm{int}} = 1.297 (\rm{A}_{1268} - \rm{A}_{2580}),
\end{equation}
where $A_{1268}$ and $A_{2580}$ are the attenuations at the wavelengths limits of the \citet{calzetti1994} windows.

The value of $n$ was varied over the range $-1.50 < n < 0.00$, in steps of 0.01, to encompass the range of $n$ values consistent with observed attenuation/extinction curves (see Fig. \ref{fig_klam}).
For each value of $n$, the sample was binned into six bins of M$_{1500}$, and the total $\chi^2$ was then computed,
\begin{equation}\label{eq_chi2_model_fit}
\chi^2 = \sum_{i} \left(\frac{\rm{model}(i) - \rm{data}(i)}{\sigma(i)}\right)^2,
\end{equation}
where the summation is over all bins and $\sigma(i)=0.1$ is taken as the intrinsic $\beta$ scatter at the faint end of the CMR estimated by \citet{rogers2014}, and is representative of the $\beta$ distributions given in Table \ref{tab_beta_values}.
The results of this fitting procedure are illustrated in Fig. \ref{fig_bc_model_cmr} and the best-fitting $n$ values and their corresponding $\chi^2$ values are listed in Table \ref{tab_bc_model_chi2}.
Figs. \ref{fig_bc_model_lf} and \ref{fig_bc_model_cmr} show data in which the nebular continuum correction as described in \ref{sec_neb_beta_effect} has been applied; however, for completeness, in Table \ref{tab_bc_model_chi2} we also report the best-fitting values without applying the nebular correction, this provides a useful comparison for the limiting case of no nebular contribution (i.e. a 100$\%$ escape fraction, and the bluest possible UV slopes).

The results show that, across all SPS models, TEx dust models with birth-cloud ages $> 5$ Myr, are ruled out at $>2\sigma$ confidence ($\chi^2 > \chi^2_{min} + 4$).
Combining this with the luminosity functions shown in Fig. \ref{fig_bc_model_lf}, we can further constrain the birth-cloud age to be $\lesssim 3$ Myr, since at ages $\geq 3$ Myr the intrinsic volume density of galaxies at $M_{1500} > -19.0$ falls below the \cite{bowler2015} $z=5$ LF.
Formally, across all SPS models, the best-fitting solutions corresponds to either the 0 or 1 Myr models, with an average best-fitting values of $n_{\rm{neb}}=-0.46$ for the nebular corrected spectra, and $n_{\rm{stellar}}=-0.59$ for the stellar continuum only models.
In general, the nebular-corrected spectra require greyer reddening slopes because the UV continuum slopes are already reddened with respect to the intrinsic stellar spectrum; however, this distinction gradually disappears at $t_{BC} \geq 10$, as the strength of the nebular continuum decreases, and the best-fitting $n$ values converge.
Finally, we find no distinction, even at the 1$\sigma$ level ($\Delta \chi^2 = 1$), between the different SPS models.
Taken all together, the results suggest a birth cloud age in the range $0 \leq t_{BC} < 3$, and a attenuation curve slope in the range $-0.6 \leq n \leq -0.4$.

It is clear that, for the TEx model, `greyer' reddening laws with Calzetti-like slopes are favoured over steeper SMC-like slopes.
From Table \ref{tab_bc_model_chi2} it can be seen that, in general, the steepness of the best-fitting reddening slope increases with increasing birth cloud age, however even at $t_{BC} = 10 - 15$ Myr the values are in the range $-0.8 \lesssim n \lesssim -1.0$, still below the SMC extinction-law value of $n_{\rm{SMC}} \approx -1.24$.
The increase in $n$ with $t_{BC}$ occurs because, as the birth cloud age increases, the intrinsic SEDs become fainter in $M_{1500}$, thus a smaller $A_{1500}$ is required to fit to the luminosity function, and a steeper $n$ is required to simultaneously redden the SEDs to match the CMR.
Although this effect is partially compensated for by the intrinsic SEDs becoming redder at larger $t_{BC}$, the compensation is not sufficient to keep the reddening law as grey as a Calzetti law.
It is possible that increasing the birth cloud age further would lead to SMC-like values of $n$, however this is really a moot point since, at $t_{BC} \geq 3$ Myr, the data become incompatible with the observed LF.
Indeed, an acceptable match to the observations can only be achieved if the UV output from stars $\lesssim$ 3 Myr old is visible in the integrated spectra of galaxies; therefore, OB star associations should be able to drift away from, or photo-evaporate, their parent molecular clouds within this timescale.
Such a scenario is consistent with the youngest ages of OB star associations observed in the Milky Way \citep[e.g.][]{massey1995}.

    \begin{figure*}
        \centerline{\includegraphics[width=9.0in]{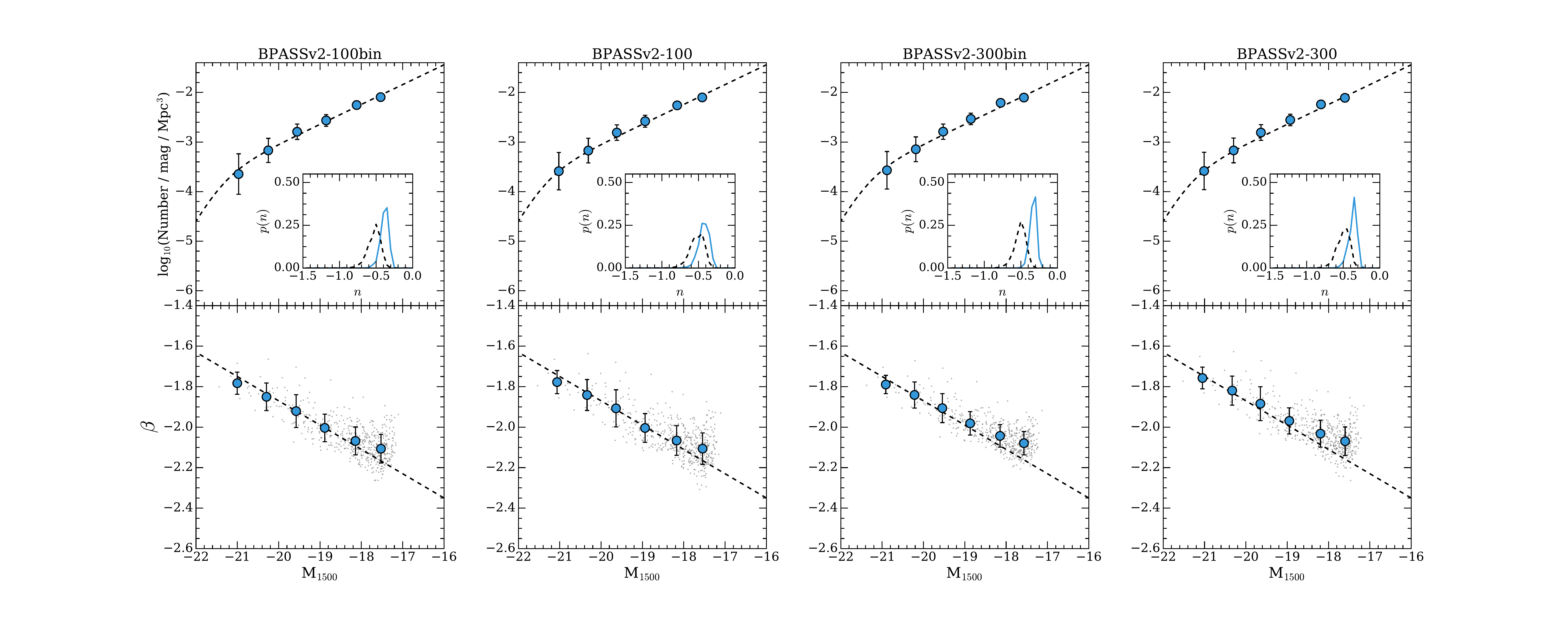}}
        \caption[]{The top panels show the $z=5$ luminosity function at the best-fitting CF2000 model solution for each of the four SPS models, compared to the luminosity function of \citet{bowler2015} (dashed black line). 
        The simulation data are shown binned into six M$_{1500}$ bins (blue circles).
        The bottom panels show the $z=5$ CMR at the best-fitting solution, with the dashed line showing the observed CMR of \citet{rogers2014}.
        Again the simulation data is shown binned into six M$_{1500}$ bins (blue circles). 
        The small grey dots in these panels are the positions of the individual galaxies. 
        In the insets in the top panels, the solid blue curves show the normalized reddening slope probability density functions, $p(n)$, derived from combining the results of a $\chi^2$ fit to the LF and CMR (see text for details).
        For reference, the dashed black line shows the probability density function assuming no nebular continuum}
        \label{fig_cf2000_cmr_lf_fit}
    \end{figure*}

In summary, comparing the results of the TEx model to the observed LF and CMR at $z=5$ leads us to conclude that: (1) no satisfactory match is achieved unless the UV output from stars $\lesssim$ 3 Myr old is visible in the integrated UV spectra of galaxies; (2) the models show no evidence for a departure from Calzetti-like attenuation laws; and (3) there is no strong preference for binary or single-star BPASSv2 models.

\subsection{CF2000}

To fit the CF2000 model we applied a more detailed analysis, performing a $\chi^2$ minimization across all parameters as listed in Table \ref{tab_cf2000_model_values}.
This allowed us to get a better estimate of the formal error on the attenuation curve slope by marginalizing of all free parameters.
The ranges of parameter values were chosen after some experimentation using coarser step sizes.
For each combination of parameter value, we calculated the optical depth as a function of wavelength for the birth cloud ($t \leq t_{BC}$) and ambient ISM ($t > t_{BC}$) as given by Equation \ref{eq_cf2000}, and reddened the spectrum accordingly.
We binned the reddened sample into six bins of M$_{1500}$ and performed a $\chi^2$ fit to both the CMR and LF.
For the fit to the luminosity function, we defined the error in each bin as the Poisson error on the number counts in that bin.
The best-fitting model parameters were determined via $\chi^2$ minimization, where the $\chi^2$ value for each set of model parameters was taken as the sum of the CMR and LF values (i.e. $\chi^2=\chi_{\rm{LF}}^2+\chi_{\rm{CMR}}^2$).
From the resulting $\chi^2$ we constructed marginalized probability density functions for each parameter.

    \begin{table}
        \centering
        \caption{The range of parameter values used in the fitting on the CF2000 model.}\label{tab_cf2000_model_values}
        \begin{tabular}{lccc}
            \hline
            Parameter & Min & Max & Step Size \\
            \hline
            \hline
            $n$ & $-1.5$ & $\phantom{-}0.0$ & 0.05 \\
            $\mu$ & $\phantom{-}0.0$ & $\phantom{-}1.0$ & 0.10 \\
            $a_{0}$ & $-4.5$ & $-1.0$ & 0.25 \\
            $a_{1}$ (d$\rm{A}_{1600}$/dM) & $\phantom{-}0.1$ & $\phantom{-}0.8$ & 0.05 \\
            $t_{BC}$ & $\phantom{-}0.0$ & $\phantom{-}15 $& 1.00 \\
            \hline
        \end{tabular}
    \end{table}

    \begin{table*}
        \centering{}
        \caption{The best-fitting $n$ values for the CF2000 model, including both the stellar + nebular continuum and stellar continuum only cases. The 1$\sigma$ confidence interval is obtained from solutions with $\Delta \chi^2=1$.}\label{tab_cf2000_fitted_values}
        \begin{tabular}{lcccccc}
            \hline
            Parameter & $n_{neb}$ & $\Delta n_{neb}$ & $\chi_{neb}^2$ & $n_{stellar}$ & $\Delta n_{stellar}$ & $\chi_{stellar}^2$ \\
            \hline
            \hline
            BPASSv2-100bin & $-0.35$  & $-0.45 < n < -0.30 $ & 1.36 & $-0.50$  & $-0.70 < n < -0.40 $ & 0.95 \\ 
            BPASSv2-100 & $-0.45$  & $-0.50 < n < -0.40 $ & 1.53 & $-0.45$  & $-0.70 < n < -0.40 $ & 1.03 \\
            BPASSv2-300bin & $-0.40$  & $-0.45 < n < -0.30 $ & 2.87 & $-0.50$  & $-0.70 < n < -0.40 $ & 0.81 \\
            BPASSv2-300 & $-0.35$  & $-0.40 < n < -0.30 $ & 2.84 & $-0.50$ & $-0.60 < n < -0.45 $ & 1.31\\
            \hline
        \end{tabular}
    \end{table*}

Fig. \ref{fig_cf2000_cmr_lf_fit} shows the LF and and CMR for the model parameters at the minimum $\chi^2$ (see Table \ref{tab_cf2000_fitted_values} for the corresponding parameter values).
The figure shows the case including nebular continuum emission but the values for the case of stellar-only SEDs (i.e. maximally blue) are also given in the table.
From Table \ref{tab_cf2000_fitted_values} it can be seen that acceptable fits can be achieved for both stellar-only and stellar $+$ nebular models.
The insets in Fig. \ref{fig_cf2000_cmr_lf_fit} show the normalized probability density functions for the slope of the reddening curve, $p(n)$.
Reassuringly, the $p(n)$ curves are consistent with the formally best-fitting $n$ values from our $\chi^2$ grid.
The formally best-fitting reddening slopes are again slightly greyer than the Calzetti law for models including a nebular continuum; however, within the estimated $1\sigma$ uncertainties of all models, the range of acceptable $n$ values is $-0.7 \leq n \leq -0.3$, consistent with both \citet{calzetti2000} and \citet{reddy2015}, but incompatible with a steep reddening law resembling the SMC extinction curve.
Furthermore, the results are fully consistent with the acceptable range of $n$ values ($-0.6 \leq n \leq -0.4$) obtained from the simpler TEx model.

\begin{figure}
        \centerline{\includegraphics[width=\columnwidth]{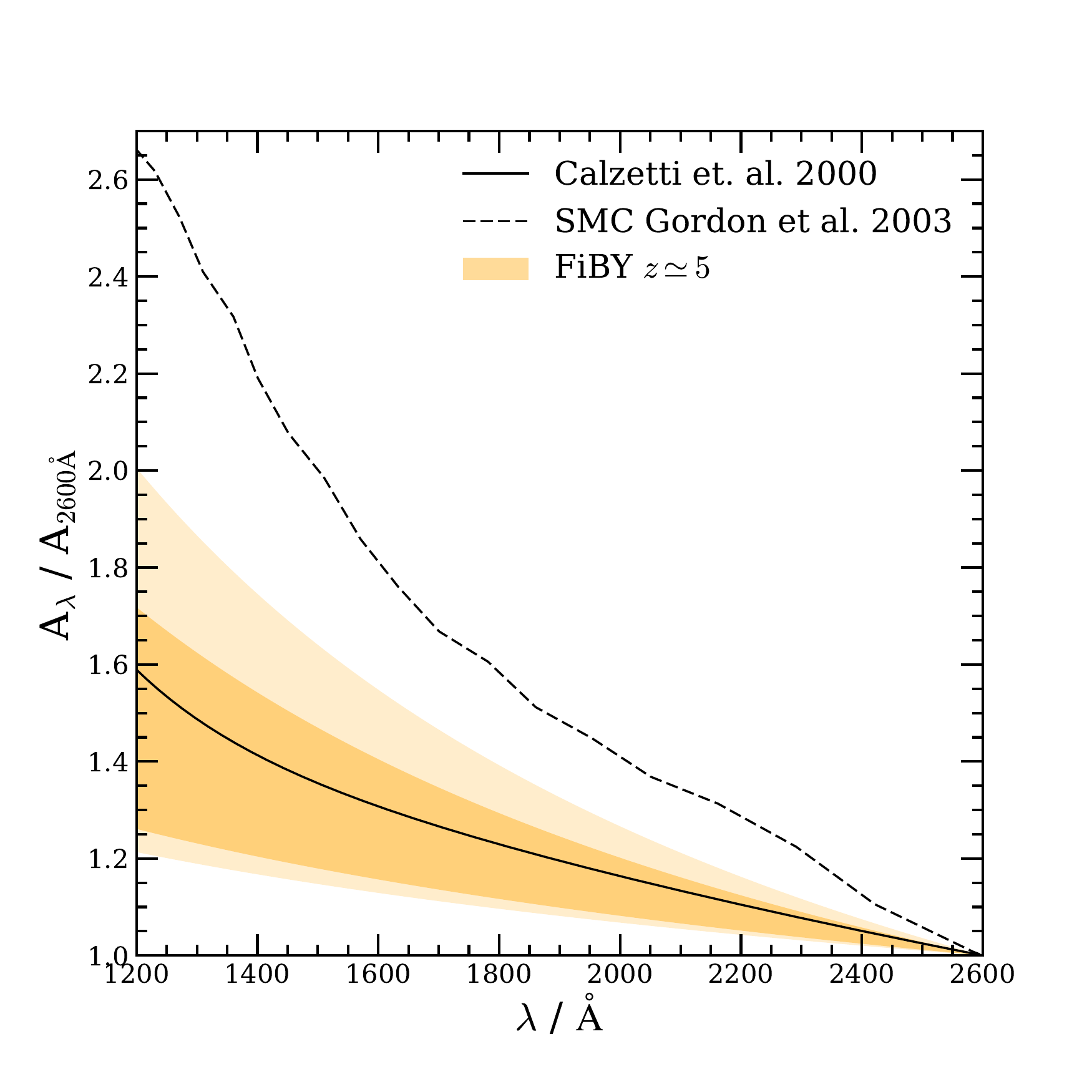}}
        \caption[]{Showing $1\sigma$ (dark orange shaded region) and $2\sigma$ (light orange shaded region) limits on the best-fitting UV dust attenuation curve from the CF2000 model.
        The black solid line shows the \citet{calzetti2000} attenuation curve and the dashed black line shows the SMC extinction curve from \citet{gordon2003}.
        All curves are normalized at $2600\rm{\AA}$.}
        \label{fig_final_attenuation_curve_plot}
    \end{figure}

Overall, across the TEx and CF2000 models, the majority of formal best-fitting values of $n$ are marginally greyer than a Calzetti law.
This is not unprecedented, since evidence for reddening curves shallower than Calzetti has been found in some direct observational studies of normal star forming galaxies at $z\simeq2$ \citep[e.g.][]{zeimann2015_dust}.
However, given the SPS and nebular continuum modelling uncertainties, it is not possible to say definitively that a shallower law is favoured by our data.
The combined analysis does, however, definitively rule out attenuation curves steeper than $n \lesssim -0.7 (-0.9)$ at $1\sigma (2\sigma)$ (see Fig \ref{fig_final_attenuation_curve_plot}). 
Crucially, we find no evidence for an attenuation curve in the UV as steep as the SMC extinction curve ($n=-1.24$), we will return to this issue in Section \ref{sec_discussion}.

\subsubsection{Other CF2000 parameters}

Although the focus of this analysis is on the slope of the reddening curve, the constraints placed on the other CF2000 model parameters were also investigated.
We found best-fitting $\mu$ values of either 0.3 or 0.4 depending on the SPS and nebular assumptions, and values within $1\sigma$ uncertainties in the range $0.1 < \mu < 0.8$.
This is consistent with the original \citet{charlot2000} results of $\mu \approx 1/3$ with a large scatter.
For the birth cloud age ($t_{BC}$) we find formally best-fitting values in the range $11-13$ Myr, however statistically-acceptable solutions at any value of $t_{BC}$ in the full parameter range ($0-15$ Myr) i.e. the birth-cloud age is not constrained by fitting to the LF and CMR within the CF2000 parameterization.
Finally, we find values of optical depth versus mass slopes (d$\rm{A}_{1600}$/dM) in the range $0.2 < \rm{d}\rm{A}_{1600}/\rm{dM} < 0.7$, with a best-fitting value of $\rm{d}\rm{A}_{1600}/\rm{dM} \simeq 0.5$.
Of course, there is a degeneracy between these parameters since they all directly affect the change in $M_{1500}$ which is mainly constrained by fitting to the LF.

\subsection{Recommended dust model parameters}

% Here we give a brief overview of the recommended parameters for both models, in the interest of any future comparisons.

% \subsubsection{TEx model}

% The extremely simple TEx model does remarkably well across all SPS models, and, given its straightforward nature, may be preferable in the sense that it can fit the data adequately with a reduced number of free parameters.
% Though the exact best-fitting solution does vary across SPS models, a birth cloud age of 1 Myr is the most commonly preferred.
% Furthermore, given that a Calzetti-like curve is fully consistent within the $1\sigma$ uncertainties of the model, we recommend the power-law index $n=-0.55$ for consistency with this commonly adopted attenuation curve.
% Fixing these two parameters, acceptable fits to the LF and CMR are found across all SPS models adopting the following relation between $\rm{A}_{1500}$ and stellar mass:
% \begin{equation}
% \rm{A}_{1500} = 0.55 \times \rm{log(M/M_{\odot})} - 3.96 + \rm{A}_{1500, BC},
% \end{equation}
% where $\rm{A}_{1500, BC}$ is the $1500\rm{\AA}$ attenuation caused by 1 Myr of complete obscuration and is $\simeq 0.1$ across all SPS models.
% This relation implies $\rm{A}_{1500}=0$ for $\rm{log(M/M_{\odot})} \leq 7.0$, and that, for an order of magnitude increase in mass, dust attenuation at $1500\rm{\AA}$ increases by 0.55 mags. 

Despite the relatively large range of parameter values over which acceptable can be found, we find that a \citet{charlot2000} type model, with parameter values not too dissimilar from the original values quoted, provides a simple and adequate framework for describing the dust attenuation law at $z \simeq 5$ from the FiBY simulation.
In the interest of future work, we provide a list of our recommend values based on the above analysis in Table \ref{tab_rec_cf200_values}.
Given that a Calzetti-like curve is fully consistent within the $1\sigma$ uncertainties of the model, we recommend the power-law index $n=-0.55$ for consistency with this commonly adopted attenuation curve.
However, as this slope is only fit to the Calzetti curve at $\lambda < 2600 \rm{\AA}$ (Fig. \ref{fig_klam}), for attenuation curves extending from UV to optical we recommend using the \citet{calzetti2000} relation explicitly or, given its similarity, the \citet{reddy2015} curve.
In addition to the values listed in Table \ref{tab_rec_cf200_values}, we recommend the following relation between $\rm{A}_{1500}$ and log(M/M$_{\odot}$), which provides the normalization of the attenuation curve:
\begin{equation}\label{eq_auv_mass}
\rm{A}_{1500} = 0.5 \times \rm{log(M/M_{\odot})} - 3.3.
\end{equation}
This relation implies $\rm{A}_{1500}=0$ for $\rm{log(M/M_{\odot})} \leq 6.6$, and that, for an order of magnitude increase in mass, dust attenuation at $1500\rm{\AA}$ increases by 0.5 mags.

% In Fig. \ref{fig_auv_mass_comparison} we show Equation \ref{eq_auv_mass} compared to the \citet{garn2010} relation for local star-forming galaxies.
% \citet{garn2010} give A$_{\rm{H}\alpha}$ as a function of log(M/M$_{\odot}$) and we have converted this to A$_{\rm{UV}}$ using the \citet{calzetti2000} law.
% Both relations are in good agreement over the mass range $8.4 \lesssim \rm{log(M/M}_{\odot}\rm{)} \lesssim 9.5$, above this mass the local relation becomes much steeper, though we caution that we suffer from low number statistics in this region and are not sensitive to heavily dust obscured star formation at the highest masses \citep[e.g.][]{dunlop2017}; at lower masses both our sample and that of \citet{garn2010} are incomplete.
% Accounting for these caveats, Fig. \ref{fig_auv_mass_comparison} implies that dust attenuation in star-forming galaxies at $z \simeq 5$ is similar to dust attenuation in the local Universe, at lease within the quoted mass range.

%     \begin{figure}
%         \centerline{\includegraphics[width=\columnwidth]{auv_lmass_comparison.pdf}}
%         \caption[]{Showing the best-fitting A$_{\rm{UV}}$ versus log(M/M$_{\odot}$) relation for our $z \simeq 5$ simulated sample (orange solid line) compared to the relation for star-forming galaxies in the local Universe taken from \citet{garn2010} (blue solid line with shaded error region).}
%         \label{fig_auv_mass_comparison}
%     \end{figure}

\section{Dust model predictions}\label{sec_IR_predictions}

In this section we discuss the infrared (IR) predictions of our best-fitting dust models, and how these compare to the latest ALMA observations at $z \simeq 5$.
Since we have derived the dust attenuation models by comparing the simulated SEDs to observed UV properties, it is useful to ask whether these same models are consistent with the, admittedly sparse, IR data at the same redshift, which trace the dust emission.
In addition, we can use our models to make predictions for future IR surveys.
The total IR luminosity (\lir) of each SED was calculated by applying the best-fitting dust model, subtracting the reddened SED from the intrinsic SED, and integrating the difference across all wavelengths i.e.:
\begin{equation}
L_{\rm{IR}} = \int_{912}^{\infty} l_{\lambda, i} - l_{\lambda} d\lambda,
\end{equation}
where $l_{\lambda, i}$ is the intrinsic SED, and $l_{\lambda}$ is the reddened SED.
This integral converges at an upper wavelength limit of $\lambda \approx 10,000 \rm{\AA}$.
Throughout this section we include the IR predictions of each SPS model for both the best-fitting TEx and CF2000 dust model, and for both the nebular continuum and stellar continuum only cases (i.e. a total of 16 models encompassing the various SED and dust prescriptions).

\subsection{ALMA HUDF $z=5$ source}

Recently, \citet{dunlop2017} (D17) have presented a 1.3-mm ALMA mosaic covering the full 4.5 arcmin$^2$ of the Hubble Ultra-Deep Field (HUDF).
Their final sample consists of 16 sources with point-source flux densities $S_{1.3} > 120 \mu$Jy at a mean redshift $\langle z \rangle = 2.15$.
Interestingly, one source in their sample (UDF12) has a spectroscopic redshift of $z=5.00$ \citep{hathi2008}, and therefore provides a useful comparison to the number counts expected over a similar area from our simulations.
UDF12 has a stellar mass log(M/M$_{\odot}$) $=9.6\pm0.12$, and a sSFR $=9.29\pm4.80$, placing it in the high mass and high sSFR tail of our sample (see Fig. \ref{fig_FiBY_galaxy_properties}).

\subsubsection{1.3mm number counts}

    \begin{figure*}
        \centerline{\includegraphics[width=8.5in]{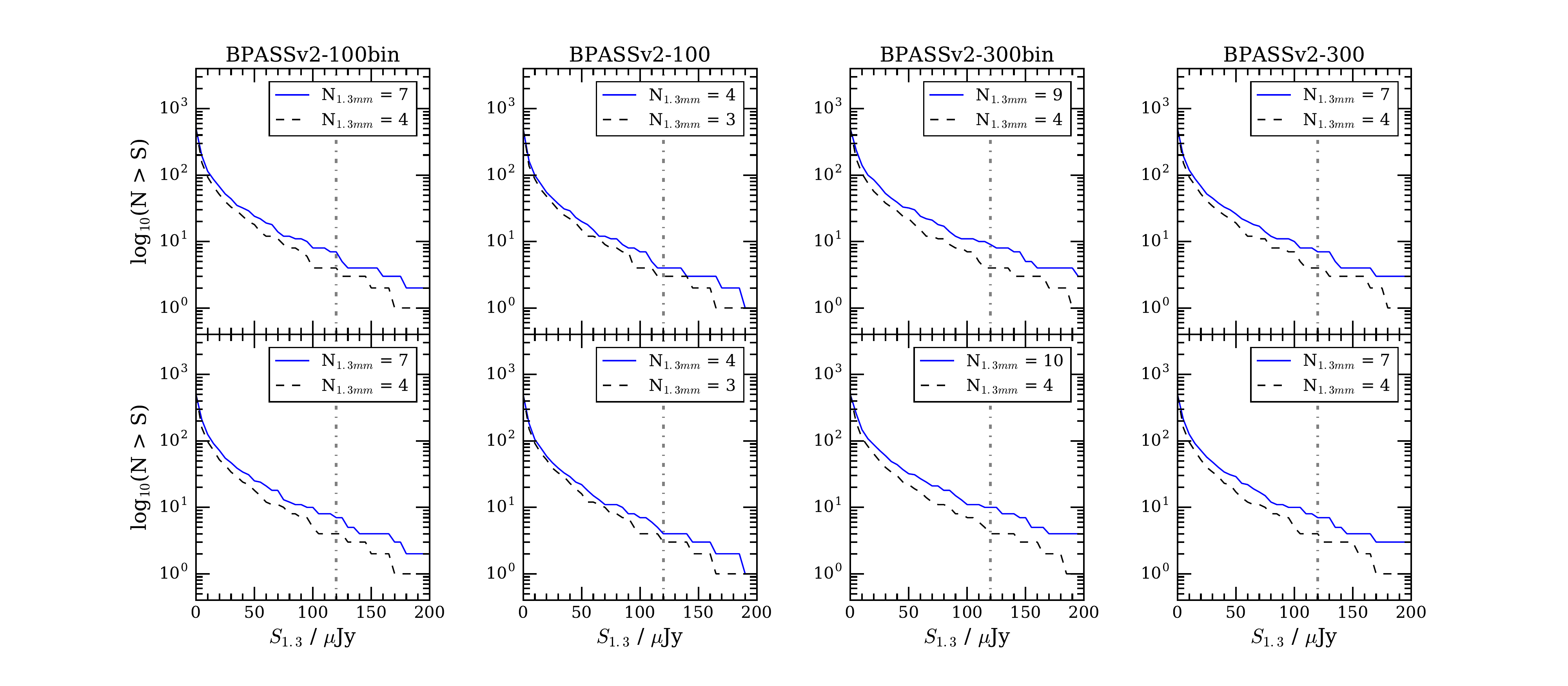}}
        \caption[]{The predicted ALMA number counts as a function of $1.3$mm flux density ($S_{1.3}$) within the $\approx 3.7 \times 10^4$ Mpc volume of the simulations. 
        Each column represents one of the four SPS models. 
        In each column the top panels show the results assuming the best-fitting parameters from the TEx model with the bottom panels showing the same for the CF2000 model. 
        In each panel the solid blue line represents the predictions for the stellar + nebular continuum SEDs and the dashed black line stellar continuum only. 
        The vertical dash-dot line in each panel is the $S_{1.3}=120 \mu$Jy flux density limit of the recent ALMA HUDF study of \citet{dunlop2017}. 
        The number of sources above this limit is quoted in the legend of each panel.}
        \label{fig_f13_ujy_counts}
    \end{figure*}

To derive $S_{1.3}$, first \lir \ was converted into a SFR$_{IR}$ using the same relation adopted by D17 \citep[from][]{murphy2011}, and $S_{1.3}$ was calculated using the conversion given in D17: $S_{1.3}$ (in $\mu$Jy) $\approx 3.3 \times$ SFR$_{IR}$ (in $M_{\odot}$yr$^{-1}$).
The number counts as a function of $S_{1.3}$ are shown by the histograms in Fig. \ref{fig_f13_ujy_counts}.
Depending on model assumptions, we find that between N $\approx 3 - 10$ galaxies have flux densities above the threshold $S_{1.3} > 120 \mu$Jy within the $\approx 3.7 \times 10^4$ Mpc$^3$ volume of our simulation, or equivalently we find a volume density of N($S_{1.3} > 120 \mu$Jy) $\approx (1-3) \times 10^{-4}$ Mpc$^{-3}$.
The number counts are larger for the models which include a nebular contribution due to the additional luminosity provided by the  nebular continuum and emission lines in these SEDs, leading to an increase in the output \lir.

Over the 4.5 arcmin$^2$ HUDF area, the comoving volume between $4.5<z<5.5$ is $\approx 1.3 \times 10^4$ Mpc$^3$, therefore, to a rough approximation, the simulation volume is $\approx 3 \times$ the HUDF volume at $z=5$.
Scaling our number counts by this factor results in a prediction of N $\approx 1 - 3$ galaxies with $S_{1.3} > 120 \mu$Jy within the HUDF ALMA mosaic, consistent with the observations.
Clearly, a comparison to number counts over such a small area of sky is hindered by cosmic variance, nevertheless these numbers provide some reassurance that we are not seriously under or over-predicting infrared flux densities.

    \begin{table}
        \centering
        \caption{Recommend values for the CF2000 model, provided to allow a simple comparison with our model.}\label{tab_rec_cf200_values}
        \begin{tabular}{lc}
            \hline
            Parameter & Value \\
            \hline
            \hline
            $n$ & $-0.55$ \\
            $\mu$ & $\phantom{-}0.30$ \\
            $t_{BC}$ & $\phantom{-}12$ Myr \\
            \hline
        \end{tabular}
    \end{table}

\subsubsection{1.3mm flux density versus UV continuum slope}

\begin{figure*}
        \centerline{\includegraphics[width=7in]{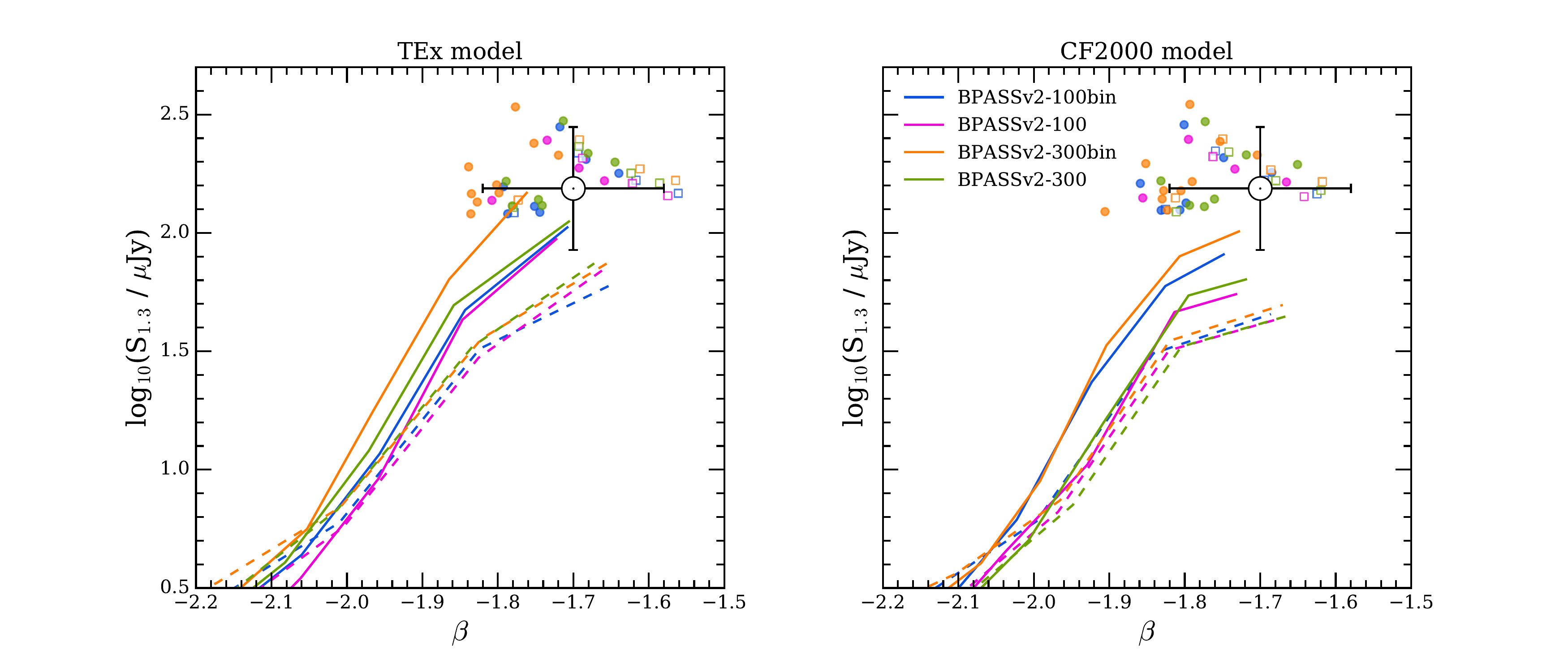}}
        \caption[]{The relationship between 1.3mm flux density ($S_{1.3}$) and UV continuum slope ($\beta$) for the TEx model (left-hand panel) and CF2000 model (right-hand panel).
        For each SPS model, the numbers are calculated at the best-fitting dust model values from the $\chi^2$ fitting described in Section \ref{sec_fitting}, and reported in Tables \ref{tab_bc_model_chi2} and \ref{tab_cf2000_fitted_values}.
        In each panel the solid lines show the running averages in six bins of $\beta$ for the four SPS models including nebular continuum (with the colour code given in legend) and the dashed lines show the corresponding stellar continuum only model.
        The small data points are the individual values for SEDs with $S_{1.3} \geq 120$ $\mu$Jy (i.e. above the sensitivity of the ALMA HUDF data) with the filled circles and open squares showing the nebular continuum and stellar continuum only cases respectively.
        The large empty circular data point with black error bars are the observed values of the one $z=5$ galaxy (UDF12) detected in the ALMA HUDF imaging \citep{dunlop2017}.}
        \label{fig_f13_versus_beta}
    \end{figure*}

In Fig. \ref{fig_f13_versus_beta} we directly compare observed quantities tracing the absolute attenuation ($S_{1.3}$) and dust law ($\beta$) of UDF12.
UDF12 has a 1.3mm flux density of $S_{1.3}=154\pm40$ \citep{dunlop2017} and a UV continuum slope of $\beta=-1.70\pm0.12$, where $\beta$ was measured by fitting a power-law to the $I_{814}$, $z_{850}$, $Y_{105}$, $J_{125}$ and $H_{160}$ photometric data points, consistent with the value obtained using a standard SED fitting procedure, in which the intrinsic SEDs are modelled as a power-laws with varying slopes  ($\beta=-1.71$).

Fig. \ref{fig_f13_versus_beta} shows that the $S_{1.3}$ and $\beta$ values extracted from the best-fitting dust models are consistent with the observation.
Although only $\approx 1 \%$ of our sample have comparable flux densities ($S_{1.3} \geq 120$ $\mu$Jy), all fall within $< 2\sigma$ of UDF12 and the majority fall within $< 1\sigma$.
Fig. \ref{fig_f13_versus_beta} also demonstrates that, for the majority of models, the flux densities of the $S_{1.3} \geq 120$ $\mu$Jy sources are biased high with respect to the average $S_{1.3}$ of galaxies with similar $\beta$ slopes.
There is no significant difference between the TEx and CF2000 models, or between the various SPS assumptions; UDF12 is consistent with the $1\sigma$ scatter on the running averages of all models.

In summary, the simple best-fitting dust models with a Calzetti-like reddening law slope ($-0.7 \leq n \leq -0.3$) derived from UV observations are simultaneously compatible with the infrared number counts and properties of sources in the ALMA HUDF image.
In other words, for at least this $z\simeq5$ ALMA source, the models account for the both the absolute attenuation and reddening across the UV-IR wavelength baseline.
However, although this is clearly encouraging, larger sample sizes at $z>5$ are needed in order to perform a more robust comparison.

\subsection{The $\bmath{z=5}$ IRX-$\bmath{\beta}$ relation}\label{sec_irx_relation}

As discussed in the introduction, the \citet{meurer1999} (M99) IRX-$\beta$ relation, and the underlying, implicit, A$_{UV}$-$\beta$ relation, has been commonly used to dust correct UV data at high redshifts, and has recently been used to argue that reddening at $z>5$ follows a SMC extinction-like curve, rather than a Calzetti curve \citep{capak2015, bouwens2016}.
In this section we explore the IRX-$\beta$, and underlying A$_{UV}$-$\beta$,  relations implied by our best-fitting models.

\subsubsection{IRX $-\beta$ relation}

\citet{meurer1999} define IRX as the ratio of far infrared flux ($F_{\rm{FIR}}$) to flux at 1600$\rm{\AA}$ ($F_{1600}$), where $F_{1600}$ is defined as a generalized flux of the form $F_{\lambda}=\lambda f_{\lambda}$, and $f_{\lambda}$ is the flux density per wavelength interval.
In their derivation of IRX$_{1600}$, they include a term (BC(FIR)$_{\rm{Dust}}$) to convert the total bolometric IR flux to a FIR flux.
In this analysis we work with luminosities, rather than fluxes, and ignore this BC(FIR)$_{\rm{Dust}}$ correction term since we can calculated the total bolometric IR luminosities of our SEDs.
In this case IRX$_{1600}$ can be written as
\begin{equation}\label{eq_irx}
\rm{IRX}_{1600} \equiv \frac{\rm{L}_{IR}}{\rm{L}_{1600}} = (10^{0.4A_{1600}}-1) \frac{\int_{912}^{\infty} l_{\lambda', i}d\lambda'}{\rm{L}_{1600,i}}.
\end{equation}
Here the first term gives the fraction of L$_{1600,i}$ absorbed by dust and $A_{1600}$ is the attenuation at $1600\rm{\AA}$.
For a given dust model, $A_{1600}$ was calculated by differencing the magnitudes at $1600\rm{\AA}$ of the intrinsic SED and the attenuated SED, where the magnitudes were calculated using a top-hat filter centered on $1600\rm{\AA}$ of width 100$\rm{\AA}$.
The second term is the ratio of the maximum amount of of luminosity available for dust heating ($l_{\lambda', i}$ is the unattenuated luminosity of the SED) divided by the intrinsic $1600\rm{\AA}$ luminosity.

Assuming the second term is a constant, this equation implies a direct transformation between the attenuation at $1600\rm{\AA}$ and IRX.
Then, since for a given attenuation curve the attenuation at $1600\rm{\AA}$ is related to the observed $\beta$ slope, the position of galaxies in the IRX-$\beta$ plane can be used to infer the properties of the attenuation curve \citep[e.g.][]{meurer1999}, although one must be wary of the crucial dependence on the intrinsic $\beta$ slopes as we will discuss below.

In \citet{meurer1999}, the second term in the equation, denoted BC(1600)$_*$, was calculated using 100 Myr duration constant star-formation rate SPS models from \citet{leitherer1995}.
They found $1.56 \leq \rm{BC}(1600)_* \leq 1.76$, adopting a final value $\rm{BC}(1600)_* = 1.66 \pm 0.15$.
Other authors have adopted the value of $1.75 \pm 0.25$ derived later by \cite{calzetti2000} \citep[e.g.][]{reddy2006,bouwens2016}.
It is worth comparing this to the BC(1600)$_*$ values of our SEDs, across all SPS models.
For the range of BPASSv2 models adopted in this paper, the median BC(1600)$_*$ across all SEDs was calculated for the stellar + nebular and stellar-only cases.
For the models including nebular continuum we find $1.70 \leq \rm{BC}(1600)_* \leq 1.78$, and for stellar continuum only models we find $1.42 \leq \rm{BC}(1600)_* \leq 1.52$.
The full range of values is consistent with original derivation of \citet{meurer1999} but emphasizes how large an effect the nebular continuum can have.
Interestingly, the $\rm{BC}(1600)_*$ values derived from SEDs including nebular continuum are in much better agreement with previous estimates \citep[e.g.][]{calzetti2000}.
For the remainder of this paper we adopt $\rm{BC}(1600)_* = 1.75$.
The IRX equation then simply becomes:
\begin{equation}\label{eq_irx}
\rm{IRX}_{1600} = 1.75 (10^{0.4A_{1600}}-1),
\end{equation}
and the relationship between IRX and $\beta$ becomes dependent only on the A$_{1600} - \beta$ relation.

\subsubsection{A$_{1600}-\beta$ relation}

If one assumes that all sources have the same intrinsic UV slopes (i.e. a constant $\beta_{i}$), then for a given reddening law slope ($n$) and a simple homogeneous star/dust mixture geometry (e.g. similar to the TEx model), there is a simple mapping between A$_{1600}$ and $\beta$ given by combining equations \ref{eq_alam_a1500} and \ref{eq_beta_vs_A_calz00}.
For a Calzetti-like reddening law ($n=-0.55$) this becomes,
\begin{equation}\label{eq_a1500_beta_calzetti}
\rm{A}_{1600, Calzetti} = 2.09 (\beta+\beta_{i}),
\end{equation}
whereas for an SMC-like law ($n=-1.24$) the slope of the relation is much shallower,
\begin{equation}\label{eq_a1500_beta_smc}
\rm{A}_{1600, SMC} = 0.99 (\beta+\beta_{i}).
\end{equation}
However, if there is an underlying correlation between $\beta_{i}$ and A$_{1600}$ (i.e. $\beta_{i}$ is not assumed to be constant across the population), then these relations do not hold.
Indeed, we find that more massive galaxies, with older UV-weighted ages, have larger values of both $\beta_{i}$ and A$_{1600}$, such that there is a general $\beta_{i} - \rm{A}_{1600}$ correlation in our sample.
This correlation is much stronger for the stellar-only models, since adding nebular continuum significantly reduces the range of $\beta_{i}$ values (see Table \ref{tab_beta_values}).
The situation is further complicated for `clumpy' geometries like CF2000, since for these models Equation \ref{eq_alam_a1500} does not apply, and so simple relations like equations \ref{eq_a1500_beta_calzetti} and \ref{eq_a1500_beta_smc} above cannot be derived.
The upshot of all this is that the slope of the A$_{1600}-\beta$ relation, derived from our models, does not simply map to the power-law slope of the reddening law as described above. 
Furthermore, the values of $\beta_{i}$ in the resulting fits are representative of the minimum values of $\beta_{i}$ of the SEDs, rather than the median value across the population.
Nevertheless, these equations are good approximations in the case of a small scatter in intrinsic UV slopes.

\subsubsection{$\rm{IRX} - \beta$ for the best-fitting dust models}

    \begin{figure}
        \centerline{\includegraphics[width=\columnwidth]{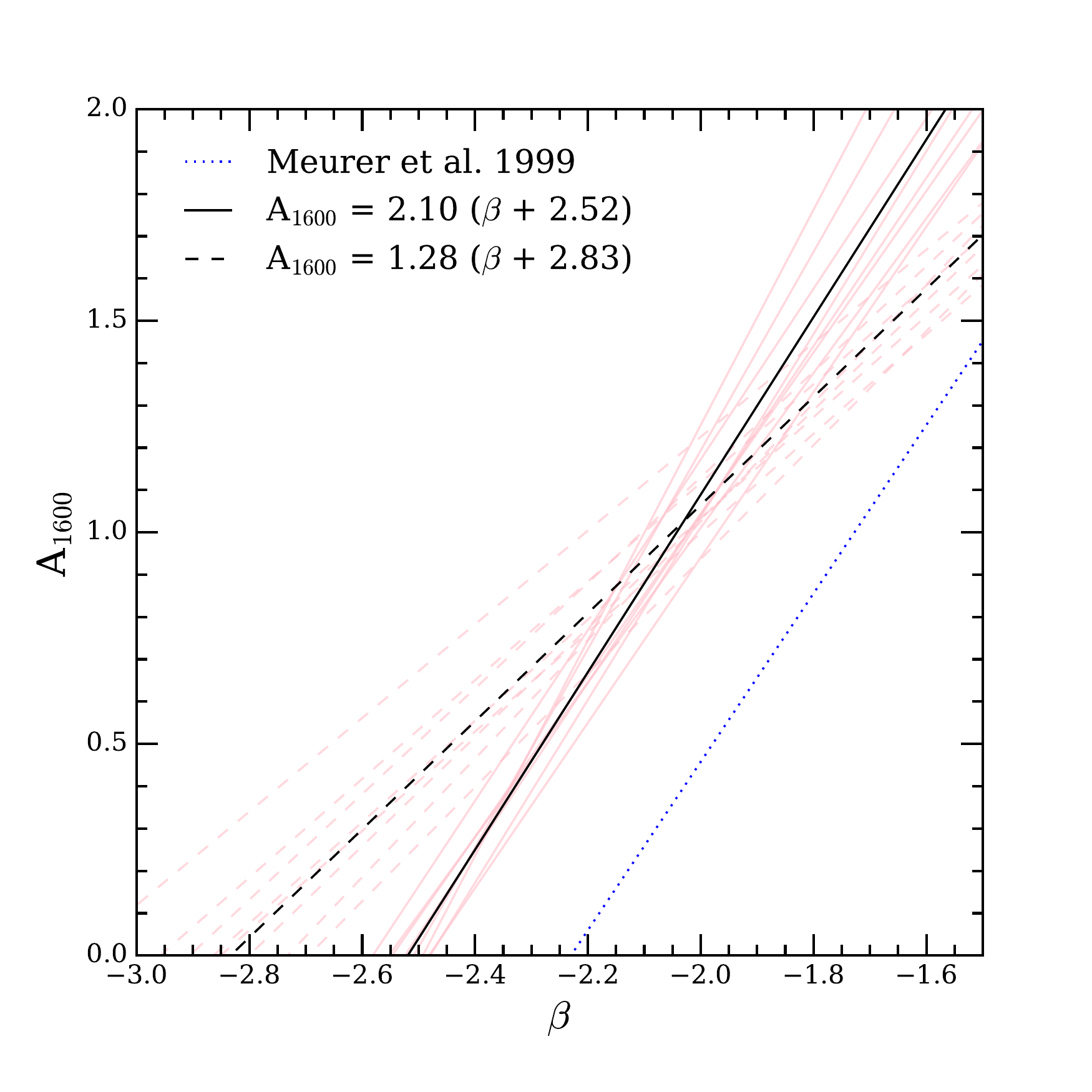}}
        \caption[]{The average $A_{1600} - \beta$ relation for both the stellar + nebular (black solid lines) and stellar-only (black dashed lines) cases.
        In each case, the $A_{1600} - \beta$ relation was derived via a linear least-squares fit to the best-fitting $A_{1600}$ and $\beta$ values combined across both dust models and all SPS models (i.e. a total of 8 individual models).
        The fits to the individual models are shown as light-pink lines for both the stellar + nebular (solid) and stellar only (dashed) cases.
        The original \citet{meurer1999} relation is shown as the blue dotted line; applying this relation to our simulated sample would underestimate the $1600\rm{\AA}$ attenuation at fixed $\beta$ by a factor $\simeq 0.6$ (and hence underestimate UV-derived SFRs by a factor $\simeq 2$).} 
        \label{fig_final_a1600_beta}
    \end{figure}

    \begin{figure}
        \centerline{\includegraphics[width=\columnwidth]{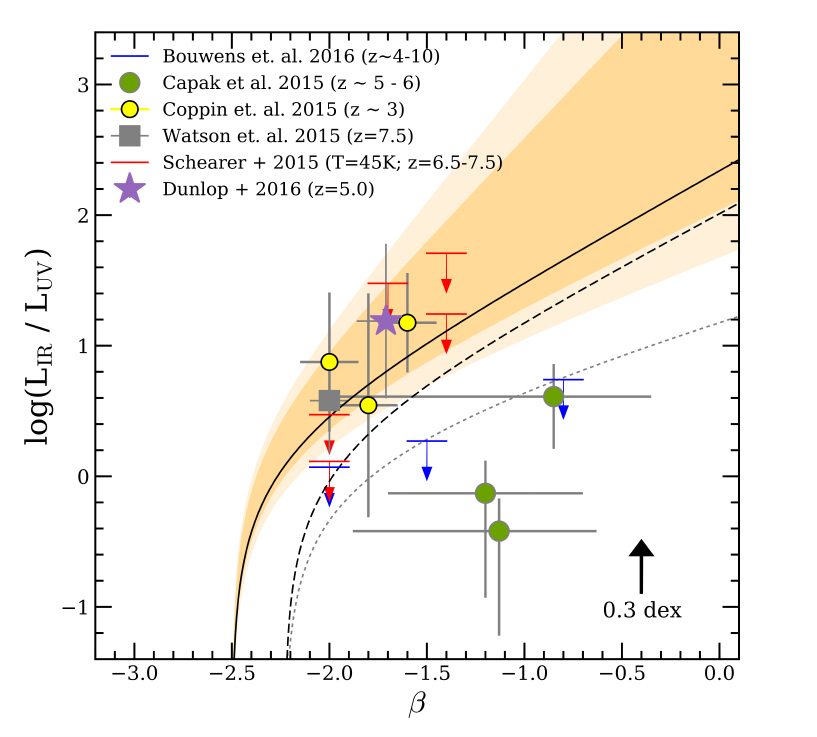}}
        \caption[]{The $\rm{IRX}-\beta$ prediction of our best-fitting model.
        The solid black line shows the predicted curve combining our best estimate of the $\rm{A}_{1600} {}- \beta$ relation (Eq. \ref{eq_a1600_beta_final}) with Eq. \ref{eq_irx}, and the shaded orange regions are the $1\sigma$ and $2\sigma$ limits assuming the limits in Eq. \ref{eq_a1600_beta_final}. 
        The dashed black line shows the original \citet{meurer1999} $\rm{IRX}-\beta$ relation and the dotted line shows the SMC relation assuming Eq. \ref{eq_a1500_beta_smc}.
        The black star shows the $z=5$ ALMA HUDF source from \citet{dunlop2017}; the grey square shows the $z=7.5$ galaxy A1689-zD1 from \citet{watson2015}; the yellow circles show the $z\simeq3$ stacks from \citet{coppin2015}; the red upper limits show the $z\simeq 6.5-7.5$ galaxies from \citet{schaerer2015} assuming a dust temperature of T=45K; the green circles show the individual detections from \citet{capak2015} and finally the blue upper limits are the stacks at $z \simeq 4-10$ from \citet{bouwens2016}; the black arrow in the bottom right-hand corner illustrates the potential upward shift of $\approx 0.3$ dex of these data assuming a higher dust temperature (see text for discussion).}
        \label{fig_final_irx_beta}
    \end{figure}

We derived A$_{1600}-\beta$ and $\rm{IRX} - \beta$ relations for our sample for both the stellar +  nebular and stellar-only cases; the results are shown in Figs. \ref{fig_final_a1600_beta} and \ref{fig_final_irx_beta}.
For each case we performed a linear least-squares fit to the best-fitting $\beta$ and A$_{1600}$ values from our models (combining the data across both dust models and all SPS models), and substituted the resulting fit into Equation \ref{eq_irx} to give the corresponding IRX $-\beta$ curve.
The best-fitting A$_{1600}-\beta$ relations are as follows (see Fig. \ref{fig_final_a1600_beta}), for the stellar + nebular case:
\begin{equation}\label{eq_neb_cont_a1600_beta}
A_{1600} = 2.10 (\beta + 2.52),
\end{equation}
and for the stellar-only case:
\begin{equation}
A_{1600} = 1.28 (\beta + 2.83),
\end{equation}
In both cases, the scatter about the relations is $\sigma=0.14$.

The relation for the stellar + nebular case is virtually identical to assuming a Calzetti attenuation law for sources with an intrinsic $\beta_{i}=-2.52$ (see Equation \ref{eq_a1500_beta_calzetti}).
In the stellar-only case, the slope of the relation is significantly shallower, this is a result of both the steeper best-fitting attenuation curves, and the stronger correlation between $\beta_{i}$ and best-fitting $A_{1600}$.
The effect of the nebular continuum is substantial, up to {}$\Delta$A$_{1600} \approx 0.5$ at the extreme ends of $\beta$ values.
The scatter of $\sigma=0.14$ around the $\rm{A}_{1600}-\beta$ relation derived above only represents the scatter about the best fitting models, to get a more robust estimate from our models we estimated $\rm{A}_{1600}-\beta$ based on the $1\sigma$ limits on $n$ ($-0.7 < n < -0.3$; see Fig. \ref{fig_final_attenuation_curve_plot} and Table \ref{tab_cf2000_model_values}).
Taking these limit into account, we estimate the following best-fitting $\rm{A}_{1600}-\beta$ relation for the stellar + nebular case,
 \begin{equation}\label{eq_a1600_beta_final}
A_{1600} = 2.10^{+1.8}_{-0.3}(\beta + 2.52).
\end{equation}
In reality, a nebular continuum contribution is expected in star-forming galaxies and, though it will likely vary for individual galaxies, we recommend Equation \ref{eq_a1600_beta_final} as the more reliable estimate of the average A$_{1600}-\beta$ relationship at $z=5$.
Futhermore, this equation is reassuringly similar to observational constrainsts based on of samples of Lyman-beak galaxies at $z \sim 3 - 6$, especially in terms of the intrinsic UV slopes at these redshits \citep[e.g.][]{debarros2014,castellano2014}.
One immediate consequence of this is that dust corrections to the UV, based on the \citet{meurer1999} IRX $-\beta$ relation, will underestimate A$_{1600}$ for an observed $\beta$, by a median of $\approx 0.6$ across the range of observed $\beta$ values $-2.5 < \beta < -1.6$, equivalent to an underestimation of the resulting star-formation rate of a factor $\approx 2$.

Finally, in Fig. \ref{fig_final_irx_beta} we show the predicted position of $z \simeq 5$ galaxies in the $\rm{IRX}-\beta$ plane along with a variety of data from the literature at $z>3$ \citep[taken from:][]{coppin2015,dunlop2017,schaerer2015,capak2015,watson2015,bouwens2016}.
It can be seen that the best-fitting line (using Eq. \ref{eq_a1600_beta_final}) is roughly consistent with the shape of the original \citet{meurer1999} curve with a shift to bluer intrinsic UV slopes of $\Delta \beta_{i} \simeq - 0.2$.
Again, this is fully consistent with typical $z\simeq5$ star-forming galaxies experiencing, on average, Calzetti-like attenuation, similar to local star-forming galaxies, yet having bluer intrinsic UV slopes based on the typically rising star-formation histories at these redshifts.
It can be seen from Fig. \ref{fig_final_irx_beta} that our predictions are in agreement with some data at high redshift \citep[e.g][]{coppin2015,watson2015}, and again are consistent with the ALMA HUDF \citep{dunlop2017}.
However, some tension clearly exists with other recent ALMA observations as we explore in more detail below.

\subsubsection{IRX-$\beta$ discrepancy with \citet{capak2015}}

It is clear from Fig. \ref{fig_final_irx_beta} that there is a significant discrepancy between the predictions of our simulations and the observed positions of the $z\approx5$ galaxies from \citet{capak2015} in the IRX $-\beta$ plane \citep[see also][]{bouwens2016}.
As discussed previously, under the assumption that the BC(1600)$_*$ factor is constant (1.75), then the position of galaxies in the IRX $-\beta$ plane is dependent, to first order, only on (i) the slope of the reddening curve and (ii) the intrinsic UV continuum slopes ($\beta_{i}$). 
Therefore, in an attempt to explain this, it is worth asking, given the above discussion on IRX $-\beta$, what type of reddening law, and what intrinsic properties of galaxies, are required to find relations compatible with the \citet{capak2015} data?

\begin{figure}
        \centerline{\includegraphics[width=\columnwidth]{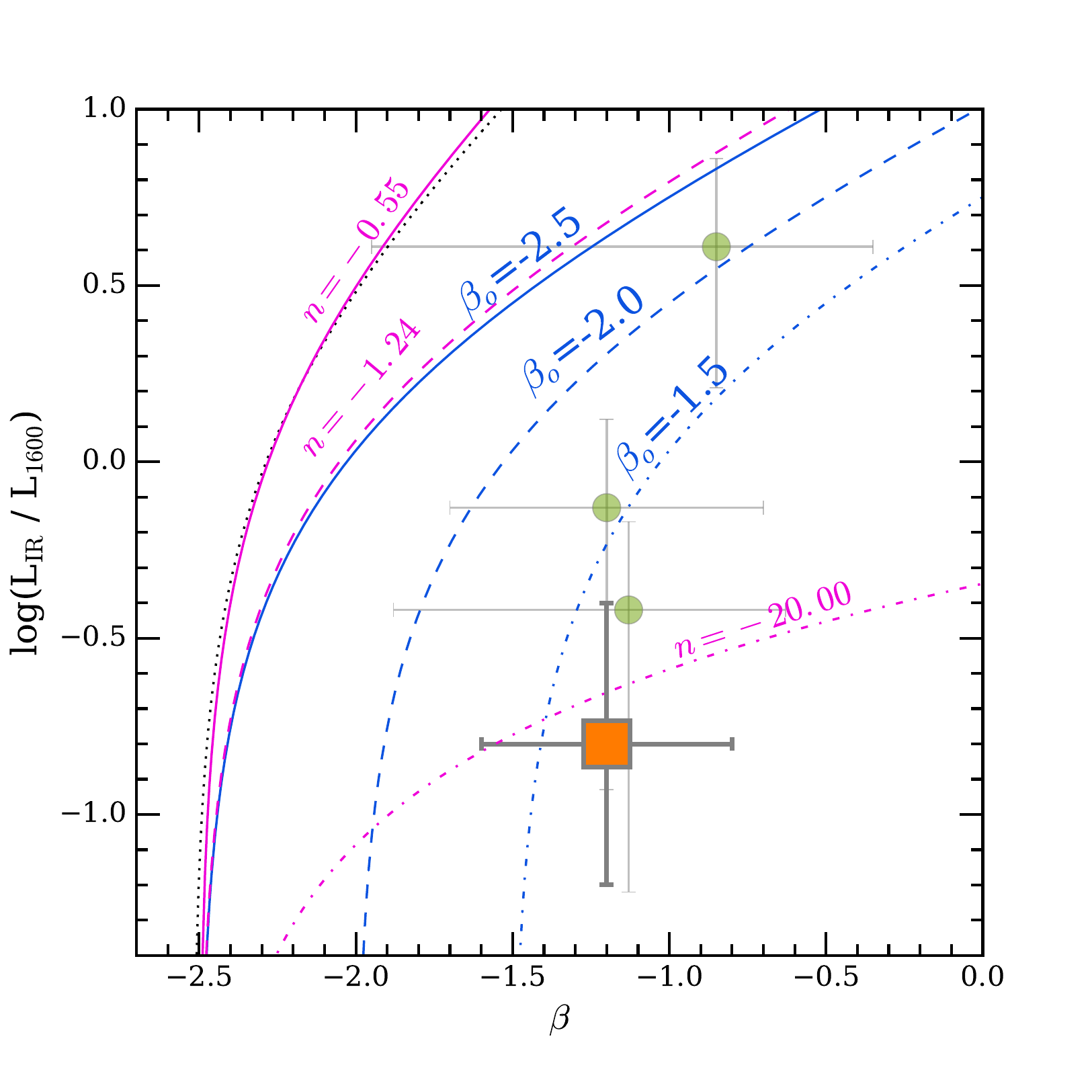}}
        \caption[]{The IRX $-\beta$ plane for a set of simple theoretical dust laws.
        The blue curves show the case for a reddening curve equivalent to the \citet{gordon2003} SMC extinction law assuming different values of $\beta_{i}$: $-2.5$, $-2.0$ and $-1.5.$
        The pink curves show the case for three simple attenuation curves of the form explored in this paper (i.e. A$(\lambda) \propto \lambda^n$) assuming different values of the exponent: $-0.55$ (Calzetti-like), $-1.24$ (SMC-like) and $-20.0$.
        As expected, the dashed pink ($n=-1.24$) and solid blue ($\beta_{i}=-2.5$) curves are compatible.
        The data points are from \citet{capak2015} as in Fig. \ref{fig_final_irx_beta}, and the black dotted line is the best-fitting IRX $-\beta$ relation from our simulations including nebular continuum (i.e Equation \ref{eq_neb_cont_a1600_beta}).}
        \label{fig_irx_beta_theory}
    \end{figure}

To do this we derived IRX $-\beta$ curves for two simple scenarios: firstly, the A$_{1600}-\beta$ relation was calculated for the \citet{gordon2003} SMC extinction curve at three values of $\beta_{i}$ ($-2.5$, $-2.0$, $-1.5$) and converted into an IRX-$\beta$ relation assuming BC(1600)$_*$ = 1.75; secondly, the A$_{1600} - \beta$ relation was calculated for a set of empirical attenuation curves of the form A$(\lambda) \propto \lambda^n$ at different values of $n$ ($-0.55$, $-1.24$, $-20.00$) and converted into an IRX-$\beta$ relation assuming a fixed $\beta_{i}=-2.5$ and BC(1600)$_*$ = 1.75.
As highlighted previously, assuming a fixed $\beta_{i}$ is a simplification, but nevertheless it is a reasonable assumption for models including nebular continuum, and adequate for the purposes of this discussion.
The derived curves are shown in Fig. \ref{fig_irx_beta_theory}, along with data points showing the three $z>5$ sources with individual detections, and the stack of undetected sources, from \citet{capak2015}.

Fig. \ref{fig_irx_beta_theory} illustrates two important points: firstly, if the reddening law at $z=5$ resembles the SMC extinction law then the implied $\beta_{i}$ of the \citet{capak2015} galaxies is $\beta_{i} \approx -1.5$.
This is possible for individual galaxies, although 2/3 of the \citet{capak2015} galaxies with ALMA detections have masses in the range probe by our simulations (log(M/M$_{\odot}$) = 9.67, 10.17, 10.39) and we do not find intrinsic slopes redder than $\beta_{i} \simeq -2.2$.
However, it is clearly incompatible with the average $\beta_{i}$ of galaxies at this redshift, since the typical \emph{observed} slopes are bluer than this value $\langle \beta \rangle \approx -2.0$ \citep{dunlop2012,dunlop2013,rogers2014}.
Secondly, if we assume $\beta_{i}=-2.5$ (i.e. the \citet{capak2015} galaxies are compatible with our simulations) then the slope of the reddening law required to explain the data is much steeper than the SMC-extinction law.
For example, the position of the stack of undetected galaxies would require $n < - 20$, much steeper than any attenuation, or even extinction law, ever observed.

Given this discrepancy, it is worth considering the potential systematics that could be affecting the \citet{capak2015} and \citet{bouwens2016} results.
On possibility is an underestimation of the dust temperature, which is taken to be 35K in both cases.
It has been suggested that dust temperature may increase with redshift due to an increase in the intensity of the radiation field heating the dust \citep[e.g.][]{bethermin2015}\footnote{This increase in radiation field intensity is also consistent with the increase in ionizing photons in star-forming galaxies inferred from the evolution of optical emission line ratios \citep[e.g.][]{steidel2014, kewley2015, cullen2016}}, and typically an upper limit of 45K has been assumed \citep{schaerer2015}.
In this case \lir \ will be underestimated by a factor $\approx 2$ and the resulting IRX underestimated by $\approx 0.3$ dex.
The magnitude of this shift is illustrated by the arrow in the bottom right-hand corner of Fig. \ref{fig_final_irx_beta}.
It would bring the \citet{capak2015} and \citet{bouwens2016} data into better agreement with our model, however it is not enough to fully account for the discrepancy.
There are other potential systematic issues affecting the determination of the UV continuum slopes, and estimated upper limits on \lir, which we will discuss in detail in a future work (McLure et al. in prep).

\subsubsection{SMC-like attenuation at lower redshifts?}

In Fig. \ref{fig_final_irx_beta} we have not included data at lower redshifts ($z \sim 2 - 3$) which also support steeper, SMC-like, attenuation laws.
In particular, \citet{reddy2010} find that galaxies with young mass-weighted ages and typically lower masses (age $\lesssim 100$ Myr; M/M$_{\odot} < 10^{10}$) have a smaller IRX, at a given $\beta$, relative to older galaxies at the same redshift.
Finally, similar conclusions have also been reached using small numbers of individual lensed galaxies \citep[e.g.][]{baker2001, siana2009}.

In \citet{reddy2010}, these young galaxies represented $\approx 13 \%$ of their full star-forming galaxy sample, and in our simulated sample $\approx 6 \%$ of galaxies have ages $< 100$ Myr.
However, since our method involves comparing to the observed luminosity function, it can only be applied to the full galaxy population complete down to a given UV magnitude, and cannot be applied to populations split by age.
Therefore, we cannot rule out the possibility that certain populations (e.g. the youngest galaxies), or any given individual galaxy, will follow a steeper attenuation curve.
We also note that, since the absolute attenuations decrease towards fainter, lower mass galaxies, the uncertainties on the form of the attenuation law will correspondingly increase, as the difference in absolute attenuation between curves becomes more difficult to distinguish.
Therefore, our conclusions are most robust at high masses.
Nevertheless, our current analysis suggests that, on average, a Calzetti-like attenuation curve is sufficient to describe the observed properties of the $z \approx 5$ galaxy population down to log(M/M$_{\odot}) \approx 7.5$.

\section{Discussion}\label{sec_discussion}

In this section, we discuss in more detail why a reddening law similar to the SMC extinction law is ruled out by our models, and how this relates to the type of dust present at high redshifts; finally we compare our results to the recent study of \citet{mancini2016}, who performed a similar analysis, but reached different conclusions.

\subsection{Why are SMC-like laws ruled out?}

It is worth asking, what differences in the intrinsic properties of the simulated SEDs are required such that a steeper attenuation law would be favored over a Calzetti-like attenuation law?
The main ways in which our simulations could be biasing our results in favour a shallower attenuation law are: (i) the intrinsic UV continuum slopes are too red; ii) the intrinsic luminosity at $1500\rm{\AA}$ is overestimated; (iii) the number density of galaxies at a given mass is overestimated; or some combination of all three.
The guiding principle here is that, for a steep law to be necessary, the change in $\beta$ for a given A$_{UV}$ needs to be larger than is currently required by intrinsic simulated galaxy properties.
Therefore, either A$_{UV}$ is kept the same and the intrinsic $\beta$ slopes must move further away from the CMR (i.e. become bluer), or the $\beta$ slopes remain the same and the UV magnitudes of the galaxies must move closer to the LF (i.e. become fainter).

\subsubsection{Steeper intrinsic UV continuum slopes?}

Firstly, if the the UV continuum slopes are too red then either the UV-weighted ages, or Fe/H values, of the galaxies are being overestimated by the simulations.
The range of UV-weighted ages of the FiBY galaxies is $20 \lesssim t_{UV} \lesssim 80$ Myr with a median of 33 Myr (see Fig. \ref{fig_beta_uvage_zfe}), and the median intrinsic stellar UV continuum slope across all models is $\beta_{i} \approx -2.6$ (see Table. \ref{tab_beta_values}).
The bluest UV slopes possible with the current BPASSv2 models are $\beta_{i} \approx -3.1$ (a starburst of age 1 Myr and metallicity 1/10 solar).
Therefore, as a simple test we fixed the intrinsic slopes of all SEDs at this value, and a $\chi^2$ fit with the TEx model was performed as described in Section \ref{sec_fitting}.
Indeed, in this case the best-fitting values of $n$ are in the range $-1.0 < n < -1.2$ (though the fits are formally significantly poorer), nevertheless it serves to demonstrate that if the intrinsic slopes were close to the steepest allowed by current SPS models, SMC-like reddening laws would become more acceptable.

\begin{figure}
        \centerline{\includegraphics[width=\columnwidth]{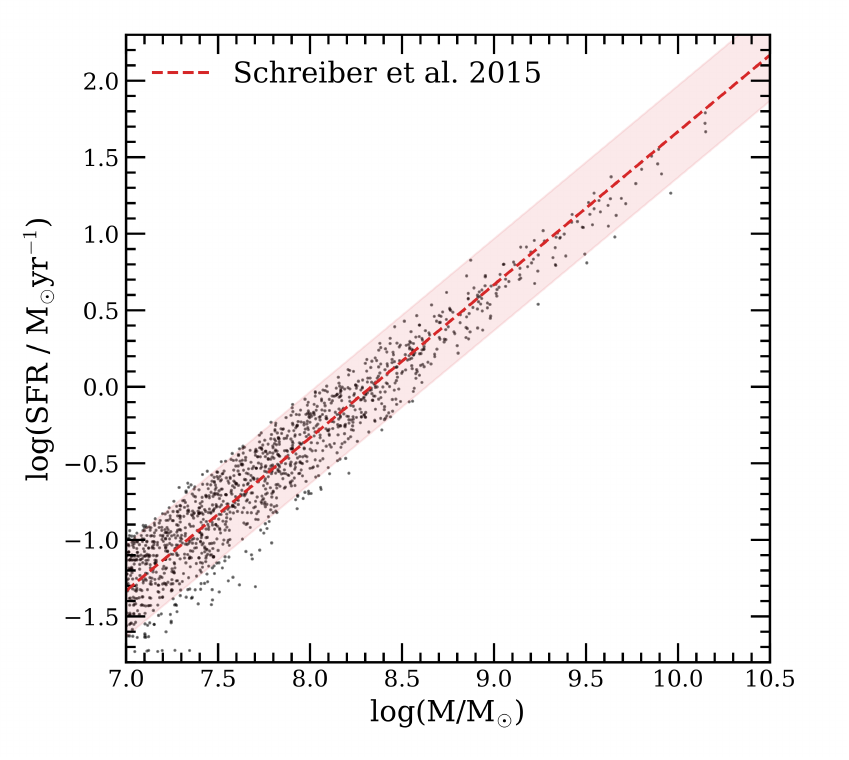}}
        \caption[]{The star-formation rate versus stellar mass relationship at $z=5$. 
        The black data points show the N=498 simulated galaxies in our sample, where the star-formation rate is averaged over the past 100 Myr and the mass is the current main-sequence mass. 
        The dashed red line shows the observed relationship taken from \citet{schreiber2015} with the shaded region showing the $\pm 0.3$ dex dispersion.}
        \label{fig_z5_main_sequence}
    \end{figure}

However, to achieve intrinsic slopes approaching this value ($\beta_{i} \lesssim -2.9$) would require that the UV-weighted ages of our galaxies be in the range $0 - 10$ Myr \citep[e.g.][]{stanway2016}.
Naturally this would lead to a boost in the SFRs and sSFRs; for example, in the extreme scenario of all stars being formed within the past 10 Myr the sSFR is 100 Gyr$^{-1}$ ($20 \times$ the median value of our sample).
However, the SFRs and sSFR of galaxies in the FiBY simulation are in good agreement with observations.
For example, Fig. \ref{fig_z5_main_sequence} compares the star-forming main sequence of our simulated sample to the observed $z=5$ relation taken from \citet{schreiber2015}, and clearly there is excellent consistency here.
Similarly, the median sSFR of our sample (4.8 Gyr$^{-1}$) is consistent with a number observations at $z\simeq5$ \citep[e.g.][]{stark2014, gonzalez2014, esther2016}.
Furthermore, as demonstrated in Fig. \ref{fig_nebular_beta_effect}, the effect of the nebular continuum increases as the intrinsic stellar continuum becomes bluer, such that the UV continuum slopes are reddened to roughly the same values $\beta_{i} \approx -2.4$.
Finally, for completeness, we briefly note that, if we have underestimated either the ionization parameter or electron density in our photoionization modelling, then we will have underestimated the contribution of the nebular continuum to the galaxy SEDs.
Assuming more extreme values of these parameters would make the intrinsic slopes redder, not bluer.
In summary, it seems unlikely that the intrinsic $\beta$ slopes at $z=5$ are being overestimated to the extent that a steep SMC-like reddening law is required to match observations.

\subsubsection{Overestimated UV luminosities?}

As we discussed previously, the intrinsic M$_{1500}$ is relatively insensitive to the metallicity and nebular continuum contribution, but rather tracks the star-formation rate over $\approx$ 100 Myr timescales.
Therefore, M$_{1500}$ will be overestimated if the SFRs of the galaxies are overestimated.
As with the UV continuum slopes, we performed a test whereby the intrinsic M$_{1500}$ of the galaxies are increased by an amount $\delta \rm{M}_{1500}$, and a $\chi^2$ fit with the TEx model carried out; we found that an increase of the order $\Delta \rm{M}_{1500} \approx 1.0$ was required to make the data consistent with a steep SMC-like reddening law.
Assuming the \citet{kennicutt2012} conversion (in which 1 M$_{\odot}$yr$^{-1}$ is equivalent to an absolute magnitude of M$_{1500} \simeq -18.75$) this shift in absolute magnitude corresponds to a factor $\simeq 2.5$ decrease in SFR, or in log space $\Delta$log(SFR)$= -0.4$, which would make the data incompatible with the observed main sequence at $z=5$ (Fig. \ref{fig_z5_main_sequence}).
Again, given the consistency of our data with observed SFRs at the same redshift, it is unlikely that we are overestimating M$_{1500}$.
Nevertheless, it is worth bearing in mind that a shift in SFR of this magnitude would be required if the observed stellar masses at these redshifts were overestimated by $\approx 0.4$ dex. 
However, given the consistency of our data with observed mass $-$ SFR relations at the same redshift, and assuming the observed masses at these redshifts are not systematically overestimated, it is unlikely that we are overestimating M$_{1500}$.

\subsection{Dust properties at $\bmath{z=5}$}

    \begin{figure}
        \centerline{\includegraphics[width=\columnwidth]{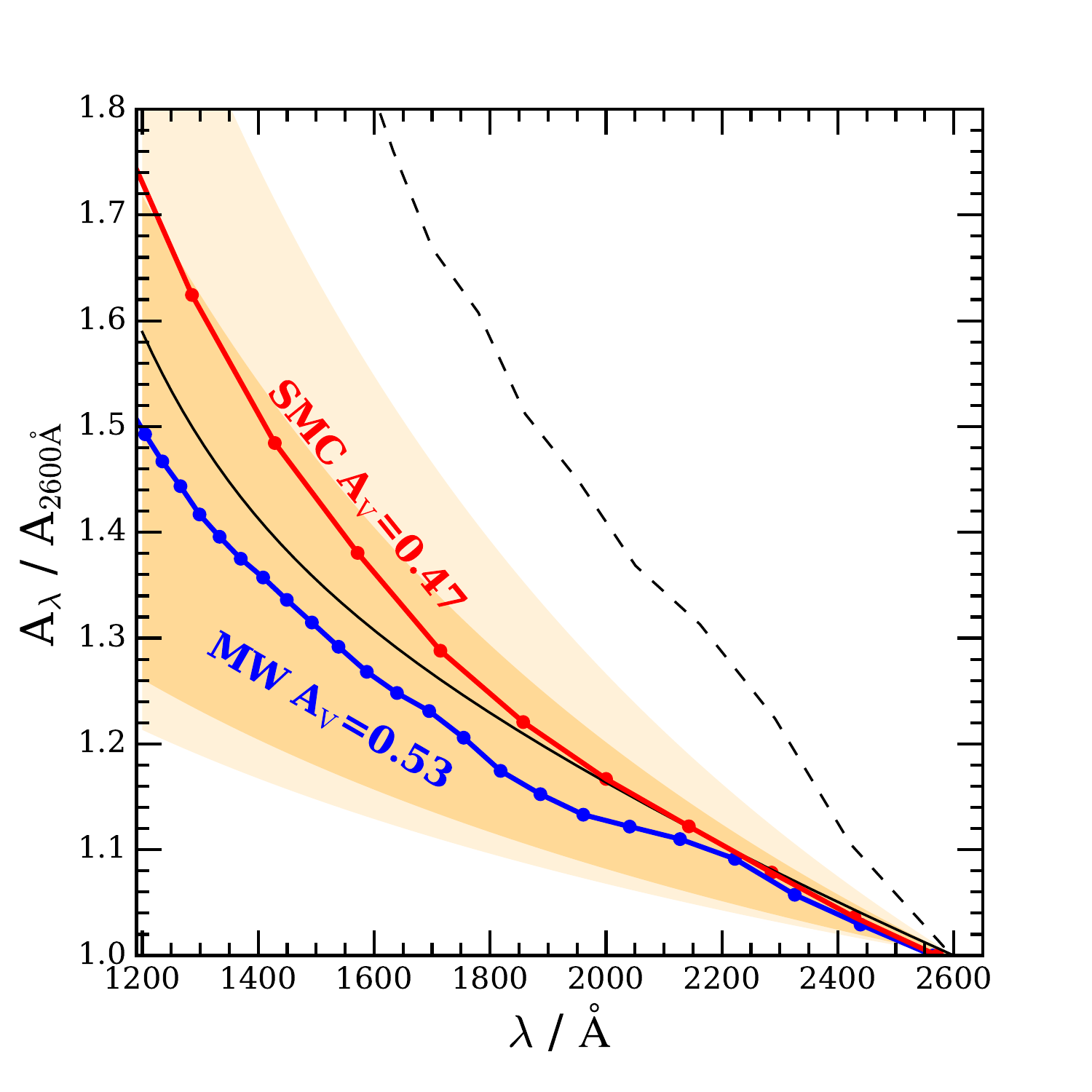}}
        \caption[]{A comparison between observed and theoretical attenuation curves at UV wavelengths.
        The black dashed and dot-dashed lines show the power-law parameterizations of the \citet{calzetti2000} and SMC (\citet{gordon2003}) curves respectively.
        The coloured lines are taken from the 3D radiative transfer models of \citet{seon2016}, where we have adopted $M_{s}=20$ and R$_{s}$/R$_{d}$=1 (see text for details).
        The red curve shows the attenuation curve resulting from the SMC extinction curve of \citet{witt2000} with A$_{V}$=0.47, and the blue curve shows the attenuation curve resulting from a MW-like extinction curve (without a $2175 \rm{\AA}$ absorption bump) with A$_{V}$=0.53, taken from \citet{weingartner2001}.}
        \label{fig_seon_curves}
    \end{figure}

What do our results mean for the `type' of dust in high redshift galaxies?
Results in the literature, which have pointed towards an attenuation curve at $z \simeq 5$ similar in shape to the extinction curve of the SMC, have been used to claim that the properties of the dust grains themselves at high redshift are more SMC-like \citep{capak2015}. 
This seems plausible, since the low metallicity of the SMC \citep[$\simeq 0.25$ Z$_{\odot}$][]{pagel1978, yozin2014}, is more representative of high-redshift galaxies, and hence the physical properties of the dust grains may also be comparable.
However, it is not necessarily true that SMC-like dust will result in an \emph{attenuation} curve similar to the SMC \emph{extinction} curve, due to the significant difference between extinction and attenuation.

This point is illustrated in Fig. \ref{fig_seon_curves}, where we show theoretical attenuation curves from the 3D radiative transfer models of \citet{seon2016}.
In these models the output attenuation curve is dependent on the system geometry (for which we have adopted the parameters  $M_{s}=20$ and R$_{s}$/R$_{d}=1$, which represent a turbulent clumpy median in which the stars are distributed throughout the total volume), the amount of dust (A$_{V}$) and the dust properties, for which we have selected, for illustrative purposes, the Milky Way model from \citet{weingartner2001}, and the SMC model of \citet{witt2000}.
The value of A$_{V}$ we have selected is typical of the median of our sample across all best-fitting models (A$_{V} \approx 0.5$).
It can be seen that, even assuming SMC dust properties, the output attenuation curve is much shallower than the SMC extinction curve, though still steeper than the Calzetti curve.
However, it is fully consistent with our best fitting model within $1\sigma$.
Assuming Milky Way dust gives better agreement with the Calzetti curve, though it is slightly shallower.
Therefore, assuming this scenario, a Calzetti-like attenuation curve at $z=5$ is consistent with dust properties somewhere intermediate between SMC and Milky Way.
Moreover, given the inherent uncertainties in radiative transfer modelling, it is not implausible to think that SMC-like dust can result in a Calzetti-like attenuation curve; indeed, as has been claimed in the past \citep[e.g.][]{witt2000}.

\subsection{Comparison to \citet{mancini2016}}

Recently, \citet{mancini2016} (M16) performed an analysis similar to the one presented here, comparing the intrinsic CMR and LF of a sample of simulated galaxies at $5 < z < 8$ to observations, in order to constrain the attenuation law and subsequent $\rm{IRX} - \beta$ relation. 
Contrary to our empirical prescription, M16 estimate the absolute attenuation (or the normalization of the reddening curve) using a semi-analytic dust evolution model which attempts to explicitly track dust growth / destruction in stars and supernovae. 
Combining this with a variety of attenuation laws and a CF2000-like model, they conclude that matching their simulations to observations re quires a steep SMC-like curve.

In the method adopted by M16, the A$_{1500}$ versus stellar mass relation is fixed by the dust model, and their relation is steeper than the one we derive by fitting to the LF (private communication).
This can be seen by inspecting Fig. 2 of M16, where the number counts at faint intrinsic magnitudes ($\rm{M_{UV}} \approx -18.0$) from their adopted simulation are much closer to the observed number counts than the number counts at faint magnitudes from FiBY (see Fig. \ref{fig_bc_model_lf} of this paper). 
In other words, M16 require much smaller dust corrections at lower masses. 
Given this, and the fact that the intrinsic UV slopes in M16 appear very similar to the median values we find in our sample ($\beta_{i} \approx −2.4$, Table 1), it is unsurprising that they require a steeper attenuation law.

One probable explanation of this discrepancy is the lower particle mass resolution of the M16 simulation (log(M/M$_{\odot}$) = 7.11 compared to 5.68 and 4.81 for the two components of the FiBY simulation) since the stellar mass range of the galaxies in our sample with $\rm{M_{UV}} \approx -18.0$ (7.6 $<$ log(M/M$_{\odot}$) $<$ 7.9) would correspond to N $\approx 3 - 10$ individual star particles in their simulation (compared to N $\approx 600 - 1000$ in FiBY\_L). 
This low resolution at faint magnitudes may result in an underestimation of the number counts (a well known effect at the resolution limit, e.g. \citet{finlator2011}), and hence lead to the perception of very little dust in low mass galaxies. 
Indeed, comparing Fig 2. of M16 to the intrinsic luminosity functions of \citet{dayal2013}, who analyse a higher resolution box from the same simulation, would appear to support this explanation. 
Without being able to perform a direct comparison to the M16 data it is difficult to identify the source of the discrepancy conclusively.
Nevertheless, we believe that the main driver is the difference in number counts at the faint end of the intrinsic LF and that this is most likely a result of the difference in particle mass resolution of the two simulations.

\section{Conclusions}\label{sec_conclusions}

We have presented an investigation of the dust attenuation law at $z=5$ using a sample of N=498 galaxies, with M$_{1500} \leq -18.0$ and $7.5<\rm{log(M/M}_{\odot}\rm{)}<10.2$, taken from the First Billion Years (FiBY) simulation.
Synthetic SEDs were generated for these galaxies using the BPASSv2 SPS models, both including and excluding the effects of massive binaries and nebular emission. 
The intrinsic UV properties of these SEDs were then compared to the observed $z=5$ LF and CMR to constrain the properties of dust attenuation.
Our main findings can be summarized as follows:
\begin{enumerate}

\item Due to the fact that the galaxies in our sample exhibit mainly rising star-formation histories, and since the age of the oldest star particles in our $z=5$ simulation ($\simeq$ 0.9 Gyr) are less than the typical timescale for the ISM to be enriched with Fe via SNe Ia ($\simeq$ 1 Gyr), we find that the galaxies in our sample have $\alpha$/Fe ratios enhanced by a factor $\approx 4$ relative to the solar value.
This is consistent with the recent observation of galaxies at $z=2.4$ by \citet{steidel2016}, and suggests care must be taken when modelling the SEDs of high-redshift galaxies, or any galaxies with a rising star-formation history, since the shape of the UV spectrum is most sensitive to the Fe/H ratio.

\item Applying a simple dust model (the TEx model) in which stars below a certain age (t$_{BC}$) are assumed to be completely obscured by their natal birth clouds, and the remaining stars experience dust attenuation assuming a homogeneous mixture of stars and dust/gas (A$(\lambda)\propto \lambda^n$), we find that the observed CMR and LF can only be recreated if t$_{BC} < 3$ Myr and $-0.6 \leq n \leq -0.3$.
The constraints on t$_{BC}$ are consistent with the age of the youngest OB associations observed in local galaxies \citep{massey1995}, and the slope of the attenuation curve consistent with a Calzetti-like attenuation curve.

\item Assuming a more complicated dust model, similar to that of \citet{charlot2000}, in which stars with ages $<$ t$_{BC}$ experience enhanced optical depths rather than being completely obscured, and again assuming $\rm{A}(\lambda)\propto \lambda^n$, we find that an acceptable match to the observed CMR and LF can only be achieved with $-0.7 \leq n \leq -0.3$.
Although other free parameters within the model (t$_{BC}$, $\mu$) are not well constrained using only these observations, the formal best-fitting values are consistent with those originally proposed in \citet{charlot2000}.

\item Across both dust models, a steep attenuation law resembling the SMC extinction law is ruled out by our simulations. 
This is in contrast to claims based on some recent ALMA observations at high redshift \citep[e.g.][]{capak2015, bouwens2016}, and the simulations of \citet{mancini2016}.
In order for the simulations to become more compatible with a steeper reddening curve, the \emph{intrinsic} UV magnitudes must be fainter and/or the \emph{intrinsic} UV continuum slopes must be steeper.
However, we argue that, based on the good agreement between the simulation data and observations of the star-forming main sequence of galaxies at $z=5$, it is unlikely that we are overestimating either of these quantities.
Nevertheless, we note that if the observed stellar masses of galaxies at $z=5$ are being overestimated by $\approx 0.4$ dex, this would result in a shift towards more SMC-like curves.

\item Comparing the IR predictions of our best-fitting models to the properties of the one $z=5$ source detected in the recent ALMA HUDF imaging of \citet{dunlop2017}, we find consistency in both the measured physical properties and implied number counts.
Nevertheless, better statistics are clearly needed to make a more robust comparison.

\item The A$_{1600}-\beta$ and resulting IRX $-\beta$ relations of our best-fitting models are inconsistent with the recent results of \citet{capak2015} and \citet{bouwens2016}, which require an attenuation curve as steep, or steeper, than the SMC extinction curve.
We find that the best-fitting IRX $-\beta$ curve is consistent with what would be expected for a simple homogeneous star/dust mixture geometry combined with the \citet{calzetti2000} attenuation law assuming all sources have intrinsic UV continuum slopes of $\beta_{i}=-2.5$.
There is clearly some tension here which can only be resolved with future, deeper, ALMA data at these redshifts.

\item Our best estimate of the $A_{1600}-\beta$ relation at $z=5$ is $A_{1600} = 2.10^{+1.8}_{-0.3}(\beta + 2.52)$; this equation implies that previous estimates of dust-corrected SFRs at these redshifts, based on the \citet{meurer1999} relation, will have underestimated the SFR by a factor $\approx 2$. 
We recommend using this relation for any young galaxy population expected to have intrinsically blue UV continuum slopes.

Overall, we find no evidence for any significant variation from the Calzetti-like attenuation curve at $z=5$.
However, more observations will be needed, both in the rest-frame UV and IR, to settle the question of the shape of the dust attenuation law at $z \geq 5$.

\end{enumerate} 

%\appendix 

\section{Acknowledgments}
FC acknowledges the support of the Science and Technology Facilities Council (STFC). 
RJM acknowledges the support of the European Research Council via the award of a Consolidator Grant (PI McLure). 
JSD acknowledges the support of the European Research Council via the award of an Advanced Grant, and the contribution of the EC FP7 SPACE project ASTRODEEP (Ref. No. 312725). 
This research made use of Astropy, a community-developed core Python package for Astronomy \citep{astropy2013}, NumPy and SciPy \citep{oliphant2007}, Matplotlib \citep{hunter2007}, {IPython} \citep{perez2007} and NASA's Astrophysics Data System Bibliographic Services.

%%%%%%%%%%%%%%%%%%%% BIBLIOGRAPHY  %% %%%%%%%%%%%%

%\begin{thebibliography}{99}
\bibliographystyle{mnras}                      % The reference style
\bibliography{fiby_z5_dust}       % Multiple bib files.
%\include{../library.bib}
%\end{thebibliography}

\label{lastpage}

\bsp

\end{document}